\documentclass[aps,prb,superscriptaddress,floatfix,twocolumn,amsmath]{revtex4-1}
\usepackage{bm,amsmath,amssymb,mathrsfs,dcolumn}
\usepackage{subfigure}
\usepackage{units}
\usepackage{hyperref}
\usepackage{mathtools}
\usepackage{xcolor}
\usepackage{wrapfig}
\usepackage{multirow}
\usepackage{tikz}
\usepackage[export]{adjustbox}
\hypersetup{pdfborder=0 0 0,colorlinks=true,citecolor=blue,linkcolor=blue}

\DeclareMathOperator*{\argmin}{arg\,min}

\definecolor{PyBlue}{rgb}{0.45094686, 0.58923722, 0.67040766}
\definecolor{PyOrange}{rgb}{1.0 , 0.75, 0.0}
\definecolor{PyBromide}{HTML}{DD5E05}
\definecolor{PyCaesium}{HTML}{33DFFF}
\definecolor{PyDeepSkyBlue}{HTML}{00BFFF}
\definecolor{PyGreen}{HTML}{008000}

\begin{document}
\title{Anharmonic lattice dynamics in large thermodynamic ensembles with machine-learning force fields:
       CsPbBr$_{3}$ a phonon-liquid with Cs rattlers}
\date{\today}

\author{Jonathan Lahnsteiner}
\email{j.lahnsteiner@utwente.nl}
\author{Menno Bokdam}
\affiliation{University of Twente, Faculty of Science and Technology and MESA+ Institute for Nanotechnology, P.O. Box 217,
7500 AE Enschede, The Netherlands}

\vspace{\fill}
\begin{abstract}
The phonon dispersion relations of crystal lattices can often be well-described with 
the harmonic approximation. However, when the potential energy landscape exhibits more 
anharmonicity, for instance, in case of a weakly bonded crystal or when the temperature is 
raised, the approximation fails to capture all crystal lattice dynamics properly. Phonon-phonon 
scattering mechanisms become important and limit the phonon lifetimes.
We take a novel approach and simulate the phonon dispersion of a complex dynamic solid at elevated 
temperatures with Machine-Learning Force Fields of near-first-principles accuracy. 
Through large-scale molecular dynamics simulations the projected velocity
autocorrelation function (PVACF) is obtained.
We apply this approach to the inorganic perovskite CsPbBr$_{3}$. 
Imaginary modes in the harmonic picture of this 
perovskite are absent in the PVACF, indicating a dynamic stabilization of the crystal. 
The anharmonic nature of the potential makes a decoupling of the system into a weakly 
interacting phonon gas impossible.
The phonon spectra of CsPbBr$_{3}$ show the characteristics of a phonon liquid. 
Rattling motions of the Cs$^{+}$ cations are studied by self-correlation
functions and are shown to be nearly dispersionless motions of the cations with a
frequency of $\sim$0.8~THz within the lead-bromide framework.

\end{abstract}

\maketitle

\section{Introduction}
First-principles based simulation methods are an important part of the present
materials science toolkit. However, most calculations are based on snapshots of
the atomic structure out of an, in some cases, very diverse thermodynamic
ensemble. Especially weakly bonded (ionic) crystals, with soft or low frequency
optical phonons, can form a pool of accessible phonon modes resulting in, for example,
non-negligible electron-phonon coupling or low thermal conductivity. 
Because of anharmonicities in the interaction potentials a harmonic spring
approximation to describe the phonons
does not suffice. With the development of machine-learning frameworks that
efficiently capture the potential energy surface described by first-principles
methods\cite{Behler:prl07,Bartok:prl10,Rupp:prl12,Bartok:prb13},
we now have the ability to explore these structural
ensembles and analyze their lattice dynamics. Previously, the computational complexity of the
density functional theory (DFT) force calculations prohibited the required length
and time scales of the simulations. We use this new capability and
simulate the phonon band structure of the CsPbBr$_{3}$, 
as presented by the projected velocity autocorrelation function (PVACF) in large-scale molecular dynamics (MD).
It serves as an example to show that this method can be applied to a much larger
class of materials, which we here refer to as "dynamic solids".
The metal-halide perovskites are materials with
technologically very attractive properties and have been under increased
research interest in recent years. They are inexpensive to produce by
crystallization from their liquid solutions\cite{Wang:nanol2018,Guo:acsel:2017},
are interesting for opto-electronic
applications\cite{Eaton:pnas2016,Chen:nanol2017,Wang:nanol2018} and for
thermoelectric applications because of their ultra-low thermal
conductivity\cite{Yuping:com2014,Mettan:jopcc2015,Filippetti:jopcc2016,Wang:nanol2018}.
The CsPbBr$_{3}$ perovskite possesses three crystallographic phases, a
low-temperature orthorhombic, a mid-temperature tetragonal and a
high-temperature cubic phase\cite{Hirotsu:jpsj1974,Jinnouchi:prl19,Sarunas:jmca2020}. Furthermore, 
it contains "rattling" Cs$^{+}$ cations locked in cavities between PbBr octahedra.

\begin{figure}[t!] \centering
	\includegraphics[width=\columnwidth]{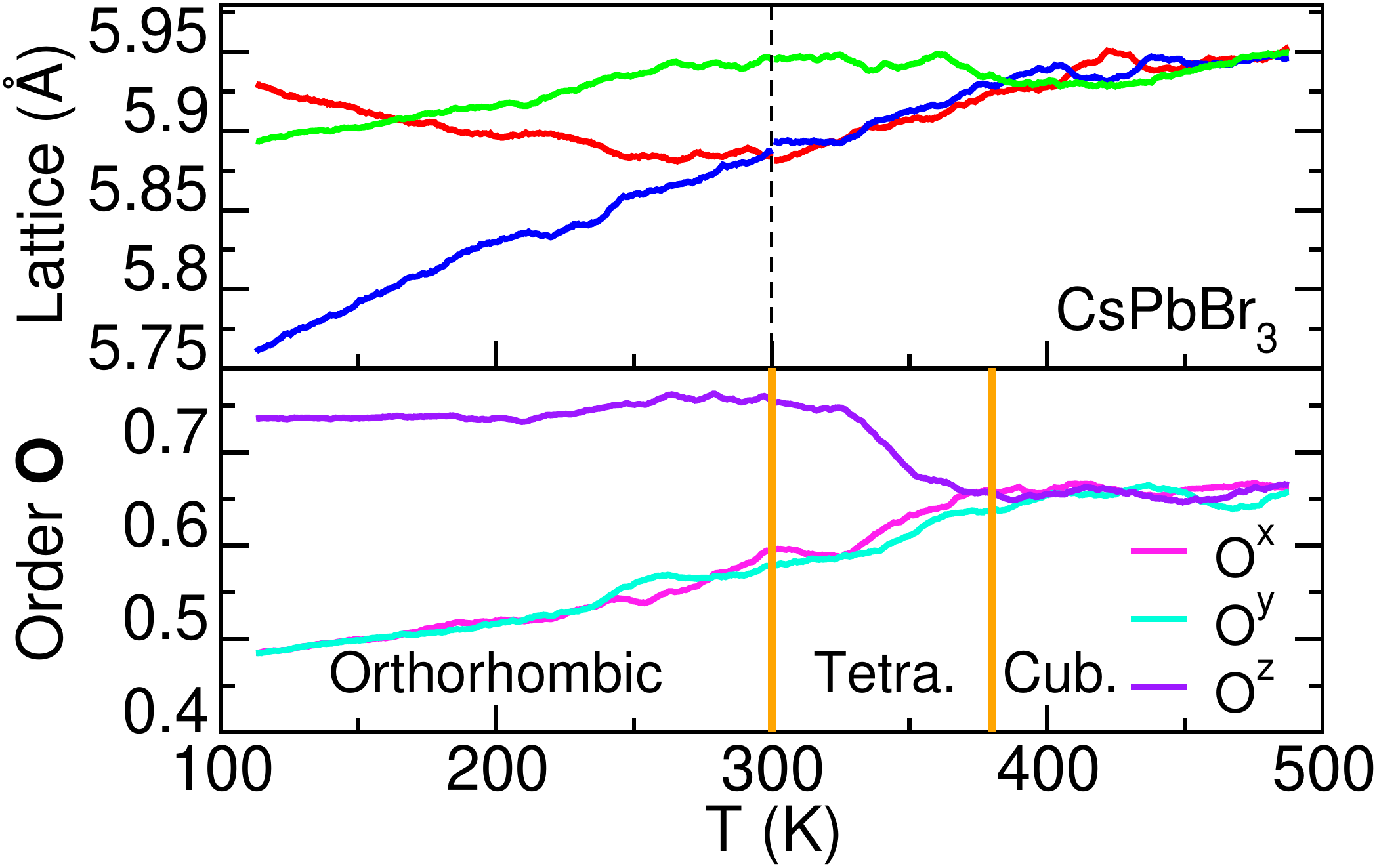}
	\includegraphics[width=1.0\columnwidth]{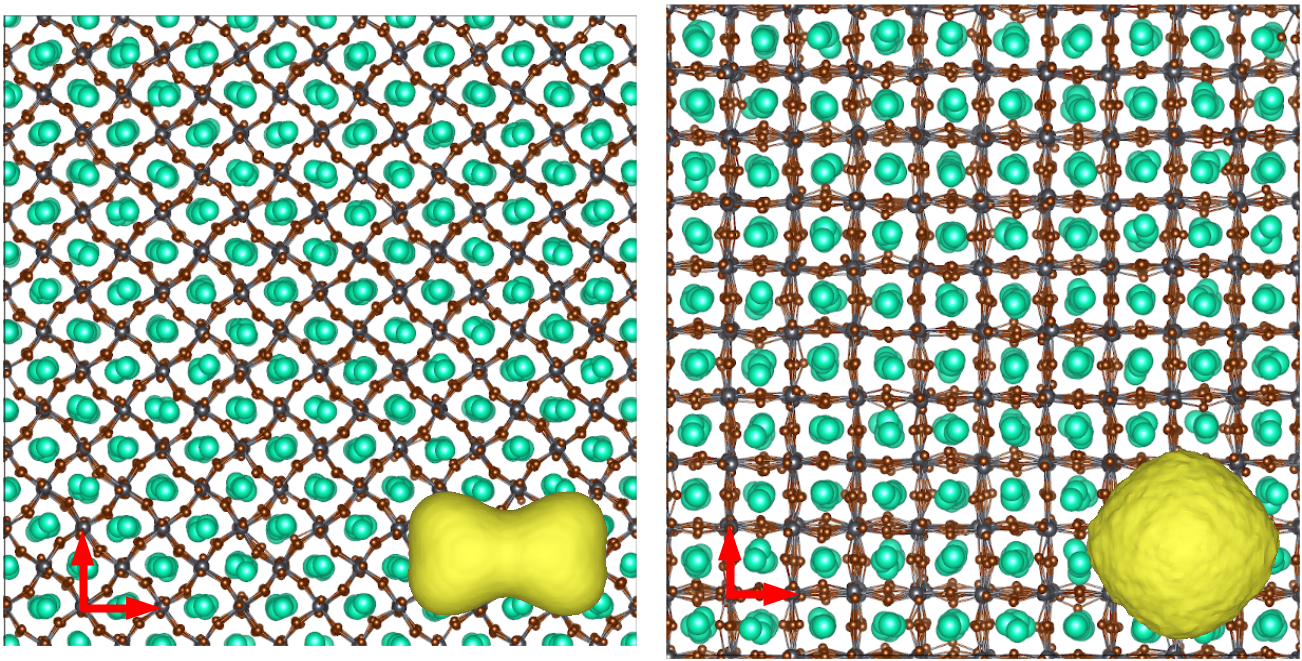}
	\caption{Lattice structure of CsPbBr$_3$ as function of temperature.
	\textit{Top}: The lattice constants in pseudo-cubic representation and PbBr
	framework order parameter vector ($\mathbf{O}$) characterize a orthorhombic
	(T~$<$~300~K), tetragonal and cubic (T~$>$~380~K) perovskite phase. Data
	obtained from slow heating $NPT$-MD with the MLFF\cite{Jinnouchi:prl19}. 
	\textit{Bottom:} Snapshots from
	the $8\times8\times8$ orthorhombic and $10\times10\times10$ cubic
	supercells during $NVT$-MD simulations at 150~K and 400~K, respectively. 
	The unitcells are indicated by the red arrows. The yellow iso-surface shows the
	displacement of the Cs from the geometric center of the PbBr cages. 
	The \href{https://homepage.univie.ac.at/menno.bokdam/sm/2021-leaf/Lahnsteiner2021-Rattling-CsPbBr3.mp4}
	{Supplementary Movie} illustrates the 'rattling' Cs$^+$ cations in the two crystal phases.
	} 
	\label{LatCsPbBr3}
\end{figure}

In this work, we study the phonon properties of this material by means of MD simulations based on recently developed \textit{on-the-fly}
trained Machine-Learning Force-Fields (MLFF)\cite{Jinnouchi:prl19}. Accurate phonon frequencies and
line widths of these materials are of great interest, because they could be used
to obtain free energies and entropic contributions to the materials
properties\cite{Sun:prb2010,Zhang:prb2014,Sun:prb2014,Zhang:prb2017}.  The
CsPbBr$_{3}$ perovksite system displays complex anharmonic dynamics with related crystal phase
transitions (see Fig.~\ref{LatCsPbBr3}). It is reported to possess closely spaced
phonon branches and overlapping line widths\cite{Simoncelli:natp19,Tadano:arxiv2021}, and highly
anharmonic coupling between the atoms\cite{Marronnier:jpcl2017,Yaffe:prl17}.  This makes
this material a very interesting candidate to study the atomic vibrations in a
framework with an accuracy beyond the harmonic approximation\cite{Chai:prb2003,Zhang:prb2014,Sun:prb2014,Guo:acsel:2017,Gehrmann:natc:19,Zhang:prb2017,Simoncelli:natp19,Hellmann:prl2020}. The phonon
properties will be studied with multiple methods: the harmonic
approximation\cite{Landau:StatPhys1,TogoPhonoPy}, the power spectrum of the
velocity autocorrelation function (VACF) and the power spectrum of the velocity
autocorrelation function projected onto the harmonic phonon eigenvectors (denoted by PVACF).

We have performed MLFF MD simulations with, for first-principles standards,
very large supercells (10,240 atoms) and long simulation times ($200$~ps per
trajectory). This enables one to resolve
the PVACF spectra on a dense reciprocal space grid. Because large supercells and
long trajectories were used to obtain converged phonon power spectra, codes for the applied
analysis methods were written. Existing tools\cite{Tadano:jpcm2014,dynaphopy,Zhang:cpc2019}
exceeded modern workstations memory requirements. In total
we used $150$~gigabytes of trajectory data per simulated crystal phase.

For CsPbBr$_{3}$ we will show the dynamic stabilization of the cubic phase, as
indicated by renormalized positive frequencies compared to the imaginary modes in
the harmonic approximation. The anharmonic coupling between the atoms and the
close-lying phonon branches induce broad and non-Lorentzian peaks in the PVACF
spectral density. Thereby the application of the weakly-interacting phonon picture to the measured
signals becomes cumbersome and non-unique.

This paper is structured as follows: In Section~\ref{CompMeth} the used
computational methods and the analysis tools are described. In
Section~\ref{Res:cspbbr} the results for CsPbBr$_3$ obtained during this study are presented.
Finally, in Section~\ref{DisConc}, the results are discussed and conclusions
are formulated.

\section{Computational Methods}\label{CompMeth}

\subsection{Machine-Learning force-field molecular dynamics} 
An \textit{on-the-fly} trained MLFF is used (in production mode, i.e. no new training)
to run MD simulations for CsPbBr$_{3}$. The generation of this force field and its
training parameters are described in detail in Ref. ~\onlinecite{Jinnouchi:prl19}. We briefly
reiterate the most physically meaningful aspects. The MLFF is trained on total 
energies, forces and stress tensors obtained from DFT.
The used algorithm automatically (on-the-fly) selects structures from the isothermal–isobaric
ensemble by a Bayes error estimate. A variant of the GAP-SOAP\cite{Bartok:prl10,Bartok:prb13} method is
used to describe the local atomic configuration for each atom. Within a cutoff radius of
6~\AA{} (radial $\rho^{(2)}(r_{ij})$) and 5~\AA{} (angular $\rho^{(3)}(r_{ij},s_{ik},\theta_{ijk})$)
probability distributions are built by adding Gaussians with a width of
0.5~\AA{}. The two-body descriptor $\rho_{i}^{(2)}$ of atom $i$ describes the probability of
finding another atom $j$ at a distance $r_{ij}$.
The three-body descriptor $\rho_{i}^{(3)}$ of atom $i$ describes the probability
to find atom $j$ at a distance $r_{ij}$ while
at the same time there is an atom $k$ at distance $s_{ik}$ from $i$, spanning an
angle $\theta_{ijk}$ between the connection vectors $\mathbf{r}_{ij}$ and
$\mathbf{s}_{ik}$. Note that in this implementation the two-body contribution
($j=k$) to $\rho_{i}^{(3)}$ is omitted, thereby creating two separable
descriptors\cite{Liu:prm21}. The obtained distributions are projected 
onto spherical Bessel functions of the order 6 and 9 for the radial and angular part, respectively.
The angular part is multiplied with spherical harmonics where the maximal
angular momentum was set to $l_{max}=6$. The coefficients of the projections
are gathered in the descriptor vector $\mathbf{X}_i$. A kernel-based
regression method is applied to map the two descriptors to a local atomic energy
\begin{equation}
U_i=F[\rho_i^{(2)},\rho_i^{(3)}]=\sum_{i_{B}}^{N_B}w_{i_B}K(\mathbf{X}_i,\mathbf{X}_{i_B}),
\end{equation} 
\noindent
where $\rho_i^{(2)}$ and $\rho_i^{(3)}$ are the total two and three body descriptors of atom i.
$N_B$ denotes the number of used local atomic reference configurations. The kernel is given by
a polynomial function
\begin{equation}
   K(\mathbf{X}_i,\mathbf{X}_{i_B})=\frac{1}{2}(\mathbf{X}^{(2)}_i
	     \cdot{}\mathbf{X}^{(2)}_{i_B})+\frac{1}{2}(\mathbf{X}^{(3)}_i\cdot{}
	        \mathbf{X}^{(3)}_{i_B})^4.
\end{equation}

On-the-fly training was performed during a $2\times2\times2$ supercell  $NPT$-MD run
in a stepwise manner. For 500~K, 370~K and 150~K, 100~ps long training runs 
were performed\cite{Jinnouchi:prl19}. The DFT
calculations use a plane-wave basis, the projector-augmented
wave method\cite{Blochl:prb94b} and the SCAN\cite{Sun:prl15} density functional
approximation. This density functional accurately describes the these types of 
lead-based halide perovskites\cite{Bokdam:prl17,Lahnsteiner:prm18}. In total 572 DFT structure 
datasets were selected by the Bayesian error
estimate during training, from which 187 (Pb), 1068 (Br) and 224 (Cs) local reference
configurations are used. A comprehensive description of the on-the-fly MLFF
generation implemented in the Vienna Ab-initio Simulation Package (VASP) is
given in Ref.~\onlinecite{Jinnouchi:prb19}.

The MLFF was shown to predict phase-transition temperatures in close agreement
with experimental observations, with predicted temperatures of $300$~K and
$380$~K for the orthorhombic to tetragonal and the tetragonal to cubic
transition, respectively\cite{Jinnouchi:prl19}. These temperatures are based on
the slow heating and cooling runs of supercells containing $6\times6\times6$
formula units. The change of the lattice parameters and change of the relative
orientation of neighbouring PbBr octahedra while heating is shown in
Fig.~\ref{LatCsPbBr3}. These findings are in close
agreement with experimental measurements based on variable temperature X-ray
diffraction analysis showing phase transitions at $361$~K and $403$~K\cite{Sarunas:jmca2020}. 
The root-mean-square errors in energy, forces and stress
between DFT and the MLFF over the total temperature interval are below
4~meV/atom, 0.05 meV/\AA{} and 1 kBar, respectively. A detailed
error analysis can be found in the supplementary material of 
Ref.~\onlinecite{Jinnouchi:prl19}. All of
the above indicates that the constructed MLFF is an
appropriate model to describe the CsPbBr$_{3}$ lattice dynamics.\newline{}

In this work, large-scale simulations of CsPbBr$_{3}$ were done at $150$~K and
$400$~K. The lattice constants for this simulations were extracted from
Jinnouchi~\textit{et.al.}\cite{Jinnouchi:prl19} and are reported in
Table~\ref{LatTable}. For the orthorhombic simulation at $150$~K a unit cell
containing 4 Pb, 4 Cs and 12 Br atoms was constructed. This unit cell was
replicated $8\times8\times8$ times, containing in total 10,240 atoms. The
$10\times10\times10$ cubic supercell was constructed from a simpler cubic unit
cell containing 5 atoms. Hence, the cubic simulation contains in total 5,000
atoms. The time steps were adjusted to $10$~fs and $5$~fs for the orthorhombic
and the cubic simulation, respectively. The two systems were equilibrated for
$100$~ps. After this initialization 20 starting structures were taken with
$100$~ps inbetween them, resulting in a total equilibration time of $2,100$~ps
and 20 molecular dynamics runs per temperature. To
obtain the trajectories required for our analysis every starting structure was
propagated for $200$~ps in the microcanonical ensemble.

\begin{table}[!h]
	\centering 
	\caption{Equilibrium unitcell lattice parameters of
        CsPbBr$_{3}$ in   the microcanonical MD simulations. Parameters adapted
	from Ref.~\onlinecite{Jinnouchi:prl19}} 
	\begin{tabular}{ | c | c | c | c | } 
		\hline 
		Temperature~[K] & $a$~[\AA{}] & $b$~[\AA{}] & $c$~[\AA{}] \\ 
		\hline 
		$150$ & 8.37 & 8.16 & 11.79 \\ 
		$400$ &	5.93 & 5.93 & 5.93  \\ 
		\hline 
        \end{tabular} 
   \label{LatTable} 
\end{table}

\subsection{Notation}\label{NotationSec}

\begin{figure}[t!] 
	\centering
	\includegraphics[width=1.0\columnwidth]{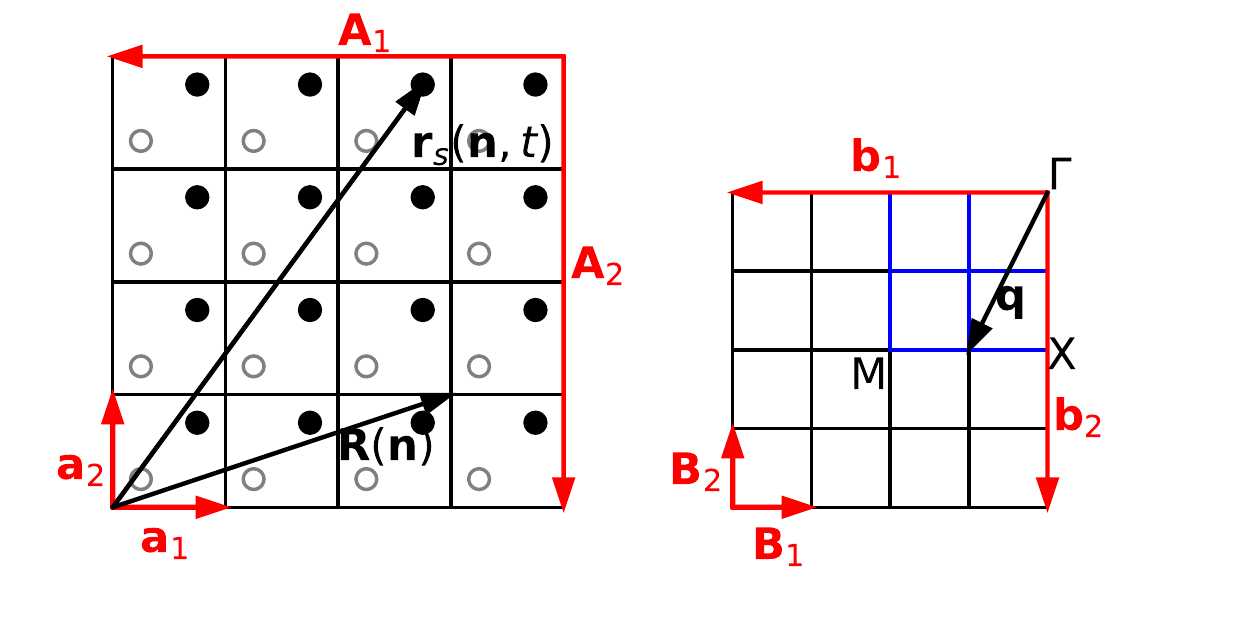}
	\caption{Left: A $4\times4\times4$ cubic primitive supercell in 2 dimensions.
		 The vectors $\mathbf{a}_{i}$ denote the unit-cell, $\mathbf{A}_{i}$
		 are the vectors defining the created supercell, 
		 $\mathbf{R}(\mathbf{n})$ denotes the unit-cell positions, and 
		 $\mathbf{r}(\mathbf{n},t)$ are the coordinates of the 
		 atoms residing in the unit-cells.\\
		 Right: Reciprocal cell of the cubic cell.
		 The vectors $\mathbf{b}_{i}$ denote the reciprocal primitive lattice.
		 The blue squares are in the first irreducible Brillouin zone of the supercell spanned by 
		 $\mathbf{A}_{i}$. The vectors $\mathbf{B}_{i}$ are the reciprocal vectors of
		 the supercell. The points named X, M and $\Gamma$ are the high symmetry points
		 of a cubic primitive lattice in the considered plane.
		 }
        \label{SketchCubicP2d}
\end{figure}

We represent crystals as supercells which
are a periodic arrangement of unit cells. A sketch of a 2 dimensional $4\times4$
supercell is shown in Fig~\ref{SketchCubicP2d} on the left hand side.
\noindent
To identify one of the unit-cells making up the supercell the vector $\mathbf{R}(\mathbf{n})$ is
used. In 3 dimensional space the vector $\mathbf{R}(\mathbf{n})$ is given by
\begin{equation}
  \mathbf{R}(\mathbf{n})=n_{1}\mathbf{a}_{1} + n_{2}\mathbf{a}_{2} + n_{3}\mathbf{a}_{3},
  \label{Harmonic1} 
\end{equation} 
\noindent 
where $\mathbf{a}_{i}$ are the lattice vectors spanning the unit-cell. The index 
vector $\mathbf{n}$ uniquely determines a unit-cell within the supercell.
The individual atoms
residing in a unit-cell can be identified by the vector $\mathbf{r}_{s}(\mathbf{n},t)$. The time 
dependence $t$ of the atomic positions comes from the molecular dynamics approach. The index 
$s$ on the vector denotes the atom index within the unit-cell. The total
number of atoms in a unit-cell is $\nu$. The lattice vectors of
the supercell are given by $\mathbf{A}_{i}=N_{i}\mathbf{a}_{i}$ and $N{i}$ is the number
of repeating unit-cells in the 3 different directions. For every supercell there exists
a reciprocal cell. An example of a reciprocal cell is shown on the right-hand side of
Fig~\ref{SketchCubicP2d}. The reciprocal unit-cell vectors $\mathbf{b}_{i}$ are 

\begin{align}
	\mathbf{b}_{1} = \frac{2\pi}
			      {\mathbf{a}_{1}(\mathbf{a}_{2}\times\mathbf{a}_{3})}
			                  \mathbf{a}_{2}\times\mathbf{a}_{3} \nonumber \\ 
			      \mathbf{b}_{2} = \frac{2\pi}
			      {\mathbf{a}_{1}(\mathbf{a}_{2}\times\mathbf{a}_{3})}
			                  \mathbf{a}_{3}\times\mathbf{a}_{1} \\ 
			      \mathbf{b}_{3} = \frac{2\pi}
			      {\mathbf{a}_{1}(\mathbf{a}_{2}\times\mathbf{a}_{3})}
			                  \mathbf{a}_{1}\times\mathbf{a}_{2} \nonumber. 
\end{align}
\noindent
The reciprocal supercell is spanned by the
vectors $\mathbf{B}_{i}=\frac{1}{N_{i}}\mathbf{b}_{i}$ where the number of repeating reciprocal unit-cells
has to match the $N_{i}$ from the real space supercell. The sampled Brillouin zone is shown in 
Fig~\ref{SketchCubicP2d} on the right hand side by the blue square. Within the first Brillouin zone
we define reciprocal wave vectors 
\begin{equation}
	\mathbf{q} = \frac{l_{1}}{N_{1}}\mathbf{b}_{1} + \frac{l_{2}}{N_{2}}\mathbf{b}_{2}+
	             \frac{l_{3}}{N_{3}}\mathbf{b}_{3},
	\label{WaveVector}
\end{equation}
\noindent
where the $l_{i}\in[0,-\frac{N_{i}}{2}]$ are a set of integers uniquely defining the accessible $\mathbf{q}$-points.
To label the wave vectors in the result sections we use fractional
coordinates $\frac{l_{i}}{N_{i}}$.

\subsection{Phonon density of states computation}\label{DOSSec}
The phonon density-of-states (DOS) $f_{s}(\omega)$,
for atom $s$ in the unit-cell is computed during an MD run from the VACF $f_{s}(t)$ given
by\cite{DoveLatBook,Sun:prb2014,Kneller:jcp2001,Dickey:pr1969}
\begin{equation} 
	f_{s}(t) =
	\frac{\left\langle\sum_{\mathbf{n}}\mathbf{v}_{s}(\mathbf{n},t')\mathbf{v}_{s}(\mathbf{n},t'')
	\right\rangle}
	{\langle\sum_{\mathbf{n}}\mathbf{v}_{s}(\mathbf{n},t'')\mathbf{v}_{s}(\mathbf{n},t'')\rangle},
        \label{PdosDyn1} 
\end{equation} 
\noindent 
where $\mathbf{v}_{s}(\mathbf{n},t')$
is the 3-dimensional velocity of an certain atom $s$ in unit cell
$\mathbf{n}$ at time t'. The time argument of the correlation function $f_{s}$ is given by $t=t'-t''$,
and the thermal average $\langle . \rangle$ is computed over the unit cells in the crystal for every atom $s$.
The phonon DOS per atom $s$ is obtained by Fourier
transforming Eq.~(\ref{PdosDyn1}) 
\begin{equation} 
	f_{s}(\omega) = \int_{\mathbb{R}}f_{s}(t)
            e^{-i\omega t}dt.  
        \label{PdosDyn2} 
\end{equation}
\noindent 
From this, the total DOS is given as the mass-weighted sum of the individual atomic
contributions 
\begin{equation} 
	F(\omega)=\sum_{s=1}^{\nu}\sqrt{m_{s}}f_{s}(\omega).
   \label{PdosDyn3} 
\end{equation} 
\noindent
The sum runs over all atoms with masses $m_{s}$ in the unit cell. 
Capital letters denote quantities reweighed by the square-root of the mass.
The phonon DOS is expected to give peaks
at all resonant frequencies $\omega$ of the studied system.
The VACF does not contain any $\mathbf{q}$ resolution and can therefore be
considered as a sum over the contributions arising from different $\mathbf{q}$.

\subsection{$\mathbf{q}$-resolved velocity autocorrelation functions}

A $\mathbf{q}$-resolved form of the VACF ($\mathbf{q}$-VACF) 
is computed analogous to Sec.~\ref{DOSSec}. This is done by Fourier transforming the mass weighed velocity field to $\mathbf{q}$-space
\begin{equation}
	\mathbf{V}_{s}(\mathbf{q},t)=\sum_{\mathbf{n}}\sqrt{m}_{s}\mathbf{v}_{s}
	      (\mathbf{n},t)e^{i\mathbf{q}\mathbf{r}_{s}(\mathbf{n},t)},
	\label{qVelocity}
\end{equation}
\noindent
Then Eq.~(\ref{qVelocity}) is self-correlated and the temporal Fourier transform is
computed
\begin{equation}
	H_{s}(\mathbf{q},\omega)=\int \mathbf{V}_{s}(\mathbf{q},t')
	     \mathbf{V}_{s}(\mathbf{-q},t'') e^{-i\omega t} dt,
	\label{Q-VACFdef}
\end{equation}
\noindent
with $t=t'-t''$ and $\text{d}t$ the corresponding differential.
The power spectrum of the so obtained function is a $\mathbf{q}$-resolved form of the 
phonon DOS~\cite{Zhang:prb2014,Sun:prb2014,Allen:prb2010,Ladd:prb1986}.

\subsection{Projected velocity autocorrelation functions}\label{PVACFMethodSec} 
The $\mathbf{q}$-VACF is decomposed by a set of phonon eigenvectors
$\mathbf{e}_{s,\alpha}(\mathbf{q})$~\cite{Sun:prb2010,Zhang:prb2014,Zhang:prb2017,Sun:prb2014,Kneller:jcp2001},
where $\alpha$ denotes the branch index. The branch index $\alpha$ goes from
1 to 3 times the number of atoms $\nu$ in the unit-cell.
Details about the definition of the phonon eigenvectors can be found in Appendix~\ref{HarmonicPhonon}.
The decomposition is done by projecting the velocities $\mathbf{v}_{s}(\mathbf{n},t)$ onto
the phonon polarization vectors $\mathbf{e}_{s,\alpha}(\mathbf{q})$
$(\in\mathbb{R}^{3})$, with PVACF in
$\mathbf{q}$ space is obtained by a spatial Fourier transform
\cite{Sun:prb2010,Zhang:prb2014,Zhang:prb2017,Sun:prb2014} 
\begin{align}
	G_{\alpha}(\mathbf{q},t') &= \sum_{\mathbf{n}}\sum_{s}\left( \left(
	                                 \sqrt{m_{s}}\mathbf{v}_{s} (\mathbf{n},t')\right)
	                                 \mathbf{e}_{s,\alpha}(\mathbf{q})\right )
					  e^{i\mathbf{q}\mathbf{r}_{s}(\mathbf{n},t')} \nonumber \\ 
				 &= \sum_{\mathbf{n}}\sum_{s}
				    \left( \mathbf{V}_{s}(\mathbf{n},t')
				     \mathbf{e}_{s,\alpha}(\mathbf{q})\right )e^{
					         i\mathbf{q}\mathbf{r}_{s}(\mathbf{n},t')}
	\label{ProjectedPhonon1} 
\end{align} 
\noindent 
The symbol
$\mathbf{V}_{s}(\mathbf{n},t)$ denotes the mass-weighted velocity
vector. The self-correlation of Equation~(\ref{ProjectedPhonon1}), 
\begin{align}
	G_{\alpha}(\mathbf{q},t) = \sum_{\mathbf{n},s,\mathbf{n}'}&
	                                         \left[\mathbf{V}_{s}(\mathbf{n},t')
 	                                         \mathbf{e}_{s,\alpha}(\mathbf{q}) \right]
	                                         \left[\mathbf{V}_{s}(\mathbf{n}',t'')
						 \mathbf{e}_{s,\alpha}(\mathbf{q}) \right] \nonumber \\ 
	                                      &\times
		e^{i\mathbf{q}\left (\mathbf{r}_{s}(\mathbf{n},t') - 
		                     \mathbf{r}_{s}(\mathbf{n}', t'') \right )},
        \label{ProjectedPhonon2}
\end{align}
\noindent
results in a VACF in $\mathbf{q}$ space projected onto the phonon
polarization vectors $\mathbf{e}_{s,\alpha}(\mathbf{q})$, with time variable $t=t'-t''$. 
A temporal Fourier transform from time $t$ to frequency
space $\omega$ is done for Eq.~(\ref{ProjectedPhonon2})
\begin{equation} 
	G_{\alpha}(\mathbf{q},\omega)=
	        \int_{\mathbb{R}}
                G_{\alpha}(\mathbf{q},t) e^{-i\omega t}dt.
        \label{ProjectedPhonon4}
\end{equation} 
\noindent 
By computing the power
spectrum $|G_{\alpha}(\mathbf{q},\omega)|^{2}$ of Eq. (\ref{ProjectedPhonon4}) we obtain the intensity of a particular
phonon eigenmode $\mathbf{e}_{\alpha,s}(\mathbf{q})$ on the
($\mathbf{q},\omega$) grid. The positions of the peaks are related to
the renormalized phonon eigen-frequencies $\tilde{\omega}_{\alpha}(\mathbf{q})$ of the
states $\mathbf{e}_{\alpha}(\mathbf{q})$.

The power spectrum $| G_{\alpha}(\mathbf{q},\omega) |^{2}$ has to show a single, well-defined
\textit{Lorentzian shaped} peak to make physical sense. This means that the eigenvector represents a phonon mode
with a particular frequency $\tilde{\omega}_{\alpha}(\mathbf{q})$ and a phonon lifetime
inversely proportional to the peak width
\cite{Zhang:prb2014,Sun:prb2014,Allen:prb2010,Ladd:prb1986}.
The eigenvectors are obtained by
a Phonopy~\cite{TogoPhonoPy} calculation. The harmonic phonon calculations were
done on $10\times10\times10$ supercells for the cubic and $8\times8\times8$ for 
the orthorhombic system. The ground state structures 
on which the harmonic approximation is computed were relaxed with an 
energy difference criterion of $10^{-4}eV$.
The FORTRAN implementations of the $\mathbf{q}$-VACF and the PVACF were added to our 
DSLEAP-code which open-source available, see Sec.~\ref{codeavail}.
The correlation functions
are computed piecewise in time such that it is possible to read the trajectory
file structure by structure. We are aware that there are several codes available
for computing the PVACF and extracting renormalized phonon frequencies such as the
DynaPhoPy\cite{dynaphopy} code, the phq \cite{Zhang:cpc2019} code or the ALAMODE~\cite{Tadano:jpcm2014}
code, giving the possibility to study anharmonic phonon frequencies.
These codes are user-friendly and handy, but they are
not applicable to the here shown simulations since they read in the whole
trajectory file at once, thereby overflowing the memory.

\subsection{Determining Cs$^{+}$ rattling frequency}\label{RattlingMethod} 
The Cs$^{+}$ cations are
locked in Pb-Br cages to which they are bound by electrostatic interactions. The
Cs$^{+}$ cations are expected to 'rattle' in their octahedra at finite
temperature. This rattling of the Cs$^{+}$ ions was proposed as one of the mechanisms responsible
for the low phonon lifetimes reported for this
material\cite{Lee:pnas17,Wang:nanol18,Simoncelli:natp19,Songvilay:prm2019}. To
determine the rattling motions in terms of a self-correlation function, the
time-dependent displacement vector $\mathbf{d}_{\text{Cs}}(t)$ of the Cs$^{+}$
cation from the geometric center (GC) of its surrounding lead framework is used.
Therefore, the first step is to determine the eight nearest lead atoms surrounding each Cs$^{+}$ ion.
Then the GC of the lead cube is computed 
$\mathbf{r}_{\text{GC}}(t)$. Thereafter, the displacement vector
$\mathbf{d}_{\text{Cs}}(t)$ is defined as 
\begin{equation}
   \mathbf{d}_{\text{Cs}}(t)=\mathbf{r}_{\text{Cs}}(t)-\mathbf{r}_{\text{GC}}(t).
   \label{rattler1} 
\end{equation}
\noindent
and its self-correlation function is given by 
\begin{equation}
	c_{Cs}(t_{1}-t_{2})=\mathbf{d}_{\text{Cs}}(t_{1})\mathbf{d}_{\text{Cs}}(t_{2}).
   \label{rattler2} 
\end{equation} 
\noindent
This results in a self-correlation function for the Cs$^{+}$ motion
relative to its GC position 
\begin{equation} 
	c_{Cs}(t_{1}-t_{2})
	= \mathbf{d}_{\text{Cs}}(t_{1})\mathbf{d}_{\text{Cs}}(t_{2}).  
	\label{rattler3} 
\end{equation}

Equation~(\ref{rattler3}) is computed for the orthorhombic and the cubic
simulation of CsPbBr$_{3}$. The obtained signals are then analyzed by means
of a Fourier transform. The power spectra of the Fourier transformed signals 
$c_{Cs}(\omega)$ are fitted by a Lorentzian function
\begin{align}
	[I,\omega_{0},\Gamma]&=
	   \argmin_{I,\omega_{0},\Gamma}\left(
	      I\left[
		\frac{\Gamma^{2}}{(\omega-\omega_{0})^{2}+\Gamma^2}\right]\right.&\nonumber\\
	&\qquad\qquad\qquad\qquad\left. - |c_{Cs}(\omega)|^{2}
            \right)^{2}.  
	\label{rattler4} 
\end{align}
\noindent
In Section~\ref{sec:results:rattle} we show that there are two separable decorrelation processes at play,
one of which we assign to rattling.

\subsection{Computation of thermal averages}
For the computation of the thermal averages we apply
two kinds of averages. First a time average over different starting
times within a single trajectory, denoted by $\langle . \rangle_{\text{time}}$ is
computed. Then the power spectra are computed and averaged over different molecular
dynamics runs, resulting in a thermal average denoted by $\langle . \rangle_{\text{Traj}}$. To 
formalize this process we will illustrate it for an example function $h_{\alpha}(\mathbf{q},t)$.
This function can be considered as any of the presented correlation functions, such as
functions ~\ref{PdosDyn1},\ref{Q-VACFdef}, \ref{ProjectedPhonon2} and \ref{rattler3}.
The time average of the correlation functions is computed by
\begin{equation}
	\langle h_{\alpha}(\mathbf{q},t) \rangle_{\text{time}}=
	     \frac{1}{N_{T}}\sum_{j=0}^{N_{T}-1}
             h_{\alpha}(\mathbf{q},t-\Delta t j), 
	\label{ThermalAverage1}
\end{equation} 
\noindent
with $\Delta t = \frac{T}{2(N_{T}-1)}$ for $N_{T}>1$. $N_{T}$ is chosen such that the 
time window after which a new starting configuration is sampled results in
$\Delta t=50$~fs for both the cubic and the orthorhombic system. This results in
200 starting configurations per trajectory.
The obtained time averaged function is Fourier transformed and the power spectrum is computed
\begin{equation}
	\left | \langle h_{\alpha}(\mathbf{q},\omega) \rangle_{\text{time}} \right|^{2} = 
			\left | \int_{-\infty}^{\infty}\langle h_{\alpha}(\mathbf{q},t) \rangle_{\text{time}}
			  e^{-i\omega t}dt \right|^{2}.
	\label{ThermalAverage2}
\end{equation}
\noindent
The power spectrum is now averaged over the 20 simulated trajectories 
per perovskite phase
\begin{align}
        \left \langle \left | \langle h_{\alpha}(\mathbf{q},\omega) 
			      \rangle_{\text{time}} \right|^{2} \right \rangle_{\text{Traj}}&=\nonumber \\
			   &\left \langle \left | \int_{-\infty}^{\infty}\langle 
			        h_{\alpha}(\mathbf{q},t) \rangle_{\text{time}}
			  e^{-i\omega t}dt \right|^{2} \right \rangle_{\text{Traj}}.
	\label{ThermalAverage3}
\end{align}
\noindent
During the analysis we also checked if the results would differ when first computing
the trajectory average $\langle \rangle_{\text{Traj}}$ and then computing the Fourier transform
and it's power spectrum. If the trajectories are long enough the order of the computation
does not matter and the two approaches result in the same spectral densities.

\subsection{Renormalized eigenfrequency determination} 
From Equation~(\ref{ProjectedPhonon4}) signals in frequency space $\omega$ are
obtained for every phonon branch $\alpha$ and $\mathbf{q}$ vector. To obtain
renormalized phonon frequencies $\tilde{\omega}_{\alpha}(\mathbf{q})$ Lorentzian
functions are fitted to Eq.~(\ref{ProjectedPhonon4}). Three
parameter Lorentzians are used, one parameter to control the height ($I$), one
parameter for the width ($\Gamma$) and a frequency parameter ($\omega_{0}$)
equivalent to the renormalized frequency
($\tilde{\omega}_{\alpha}(\mathbf{q})$). The fitting procedure is described by,
\begin{align}
	[I_{\alpha}(\mathbf{q}),\tilde{\omega}_{\alpha}(\mathbf{q}),\Gamma_{\alpha}(\mathbf{q})]&=
	   \argmin_{I,\omega_{0},\Gamma}\left(
	      I\left[
		\frac{\Gamma^{2}}{(\omega-\omega_{0})^{2}+\Gamma^2}\right]\right.&\nonumber\\
	&\qquad\qquad\qquad\qquad\left. -|\langle G_{\alpha}(\mathbf{q},\omega)
	    \rangle|^{2}\right)^{2}.  
	\label{LorentzianThefirst} 
\end{align} 
\noindent 
The three parameter Lorentzian is fitted with a linear least-square error to the
signals $|\langle G_{\alpha}(\mathbf{q},\omega)\rangle|^{2}$. The parameter $\Gamma$
is related to the full width at half maximum (FWHM) by
$(\text{FWHM)}_{\alpha}=2\Gamma_{\alpha}$ from which the phonon lifetimes can be computed
by \cite{Zhang:prb2017}
\begin{equation}
	\tau_{\alpha}(\mathbf{q})=\frac{1}{\left(\text{FWHM}\right)_{\alpha}(\mathbf{q})}=
	                 \frac{1}{2\Gamma_{\alpha}(\mathbf{q})}. 
	\label{PhononLifeTimes}
\end{equation}

\section{Results}
\label{Res:cspbbr}


\subsection{Cs$^{+}$ Rattling in CsPbBr$_3$}
\label{sec:results:rattle}
\begin{figure}[b!] 
        \centering
        \includegraphics[width=1.0\linewidth]{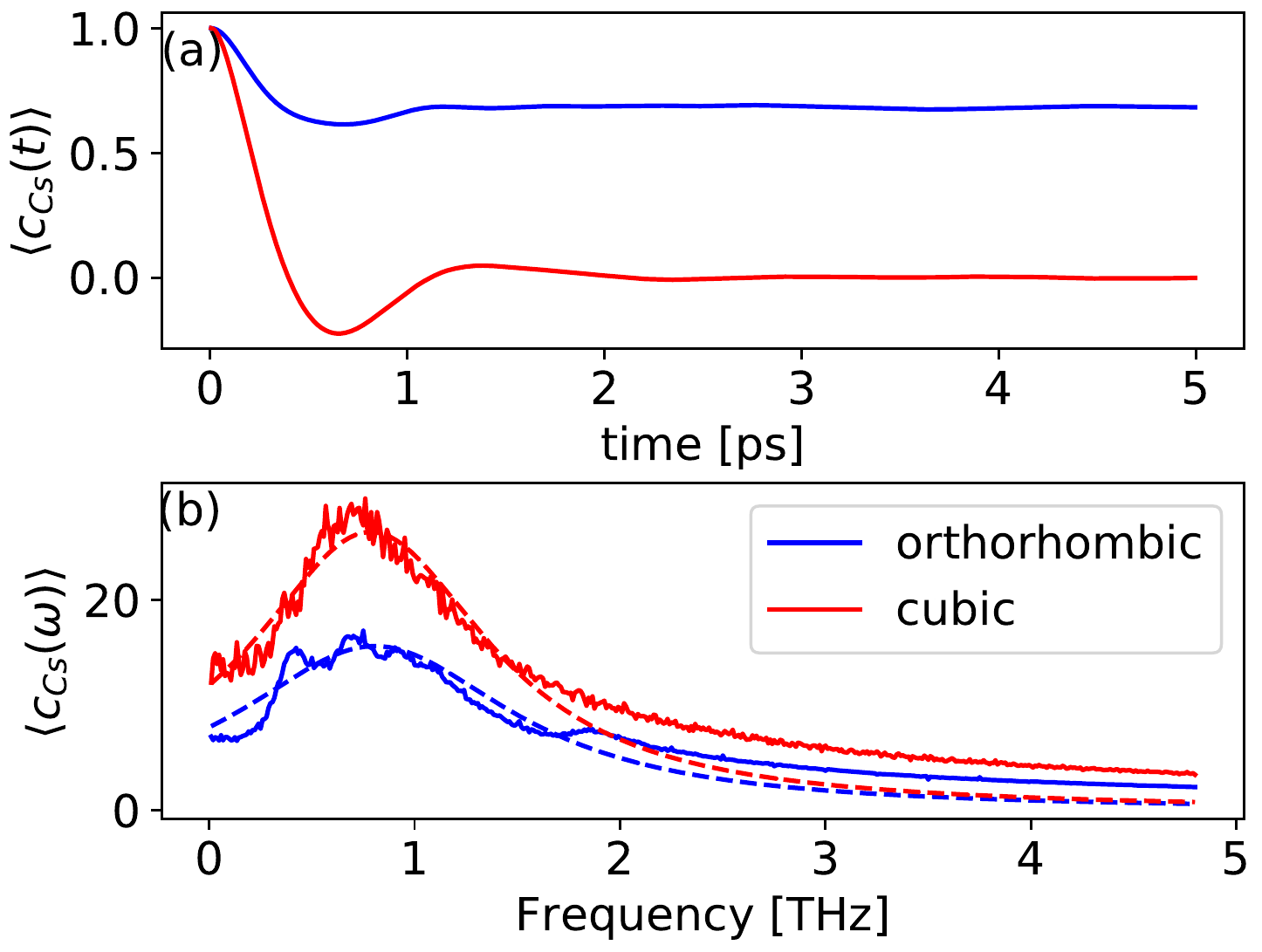}
        \caption{Averaged self-correlation function of the displacement vector
        connecting the Cs atom to the geometric center of the surrounding PbBr
        cage.} 
        \label{DynStrucFig12}
\end{figure}

Figure~\ref{LatCsPbBr3} shows that every $Cs^{+}$ cation in the CsPbBr$_3$ perovskite is surrounded by a lead bromide
framework. Since the $Cs^{+}$ cations are only weakly bound to the framework
by electrostatic interactions the $Cs^{+}$ ions will undergo rattling
motions at finite temperature. The exact nature of these dynamics is unclear.
We study the rattling motions of the $Cs^{+}$ cations by analysing the local order
parameter described in section \ref{RattlingMethod}.
The Cs rattlers scattering with phonons that propagate through the crystal 
might be a possible explanation for the low phonon lifetimes \cite{Lee:pnas17,Wang:nanol18}.
We have calculated the displacement
vector, see Eq.~(\ref{rattler1}), of the $Cs^{+}$ with respect to the (geometric center) GC
of the surrounding eight Pb atoms. Iso-surfaces of the 3-dimensional distributions of these
vectors are shown in yellow color in Fig.~\ref{LatCsPbBr3} (bottom). 
Its self-correlation function is computed
and is shown in Figure~\ref{DynStrucFig12}(a) for the cubic and orthorhombic phase.
In the orthorhombic phase, the function converges to a plateau above zero.
This shows that the $Cs^{+}$ cations at 150~K rattle around a fixed point
\textit{away} from the GC. At 400~K the $Cs^{+}$ displacement decorrelates
completely, indicating that the $Cs^{+}$ rattles \textit{around} the GC.

Figure~\ref{DynStrucFig12}(b) shows the Fourier transforms of the time
signals. There are two underlying decorrelation processes: First, there is a random thermal motion 
(Brownian-like) of the Cs$^{+}$ around its
actual position described by an exponential function $e^{-kt}$\cite{Kubo:StatPhys2}.
Secondly, a repositioning within the occupied lead cube to which we assign 
the rattling dynamics, described by $\text{cos}(\omega t)$.
The related frequencies were estimated by fitting the results in the time domain (Eq. \ref{rattler3})
of the cubic $400$~K simulation with $e^{-kt}\text{cos}(\omega t)$.

In the frequency domain, the parameter $k$ of the exponential is related to the width $\Gamma$ of the 
Lorentzian in Equation (\ref{rattler4}), and $\omega$ of the cosine is related to $\omega_{0}$.
Therefore, the position $\omega_{0}$ of the Lorentzian describes the rattling frequency of
the Cs$^{+}$ cations. 

The fitted function shown by the dashed lines in Fig.~\ref{DynStrucFig12}(b)
assigns very similar rattling frequencies of $0.81$~THz and $0.79$~THz for the orthorhombic and
the cubic phase, respectively. This shows that the rattling period is $\sim{}1.2$~ps. There
also is a smaller peak in the orthorhombic phase at 1.9~THz. This rattling frequency is not visible in the cubic phase. 
The random thermal motion is faster, roughly $2.5$~THz, and similar in
both phases. To illustrate the rattling motions two movies for the orthorhombic and the cubic 
phase were created and added as \href{http://www.dynamicsolids.net/sm/2021-leaf/}{Supplementary Movie}.
For a better visualization of the rattling motions, a period of 0.4~ps has been set as
the window size for a running average over two $NVE$ trajectories of 100~ps.
In the following sections we will study the lattice dynamics of CsPbBr$_3$ and attempt to decompose
them by means of a set of phonon eigenstates.

\begin{figure*}[t!] 
	\centering
	\includegraphics[width=1.0\columnwidth]{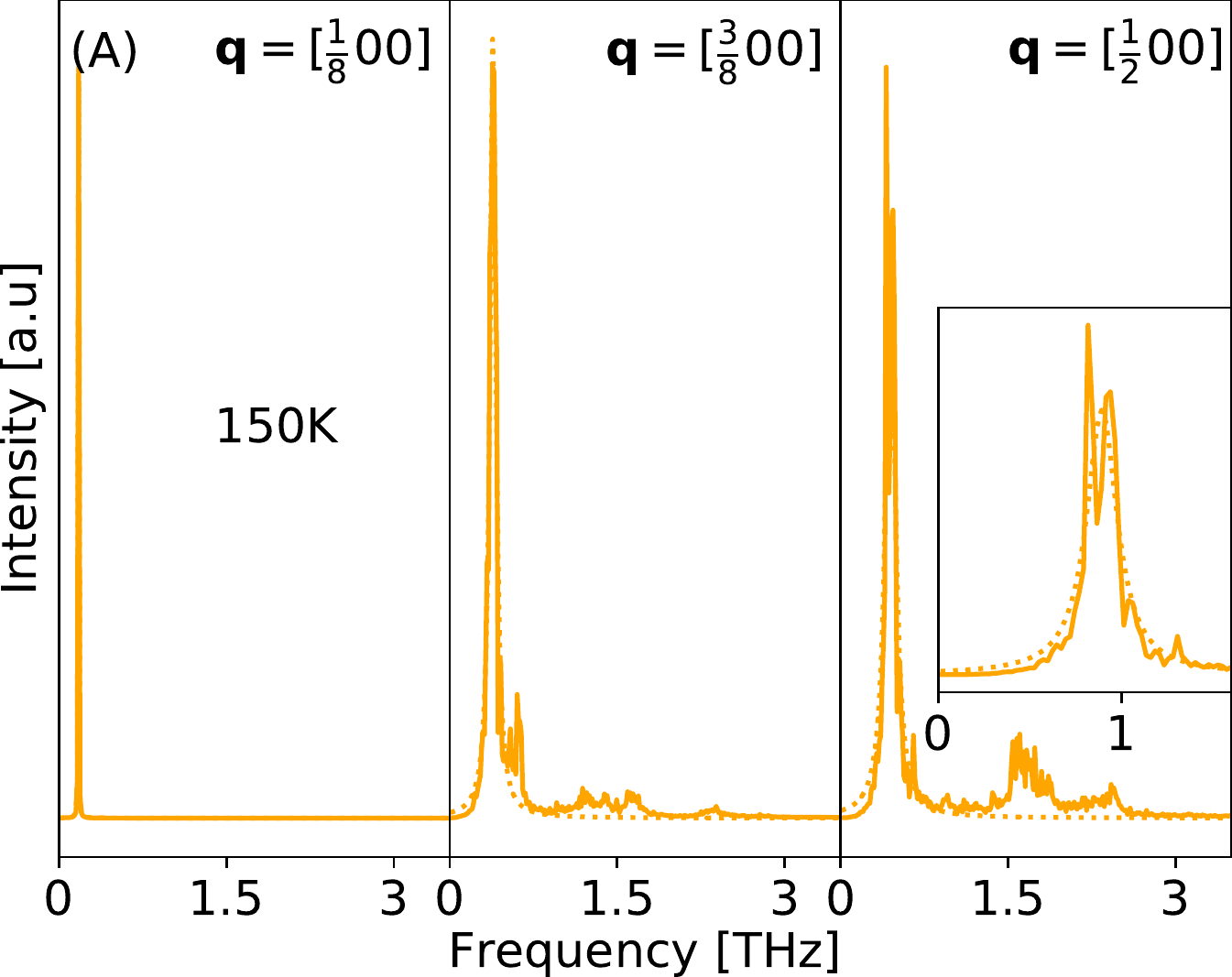}
	\includegraphics[width=1.0\columnwidth]{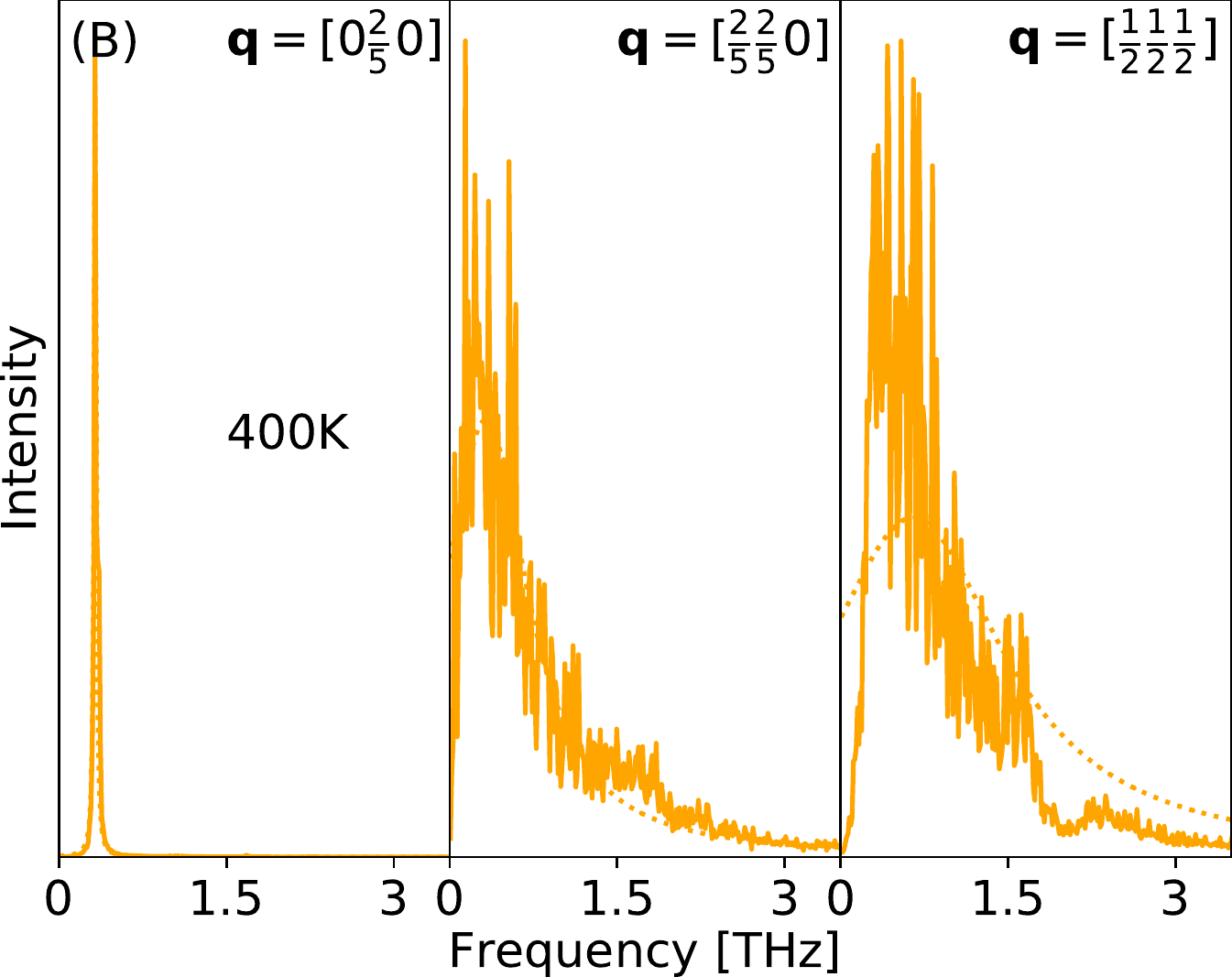}
	\includegraphics[width=1.0\columnwidth]{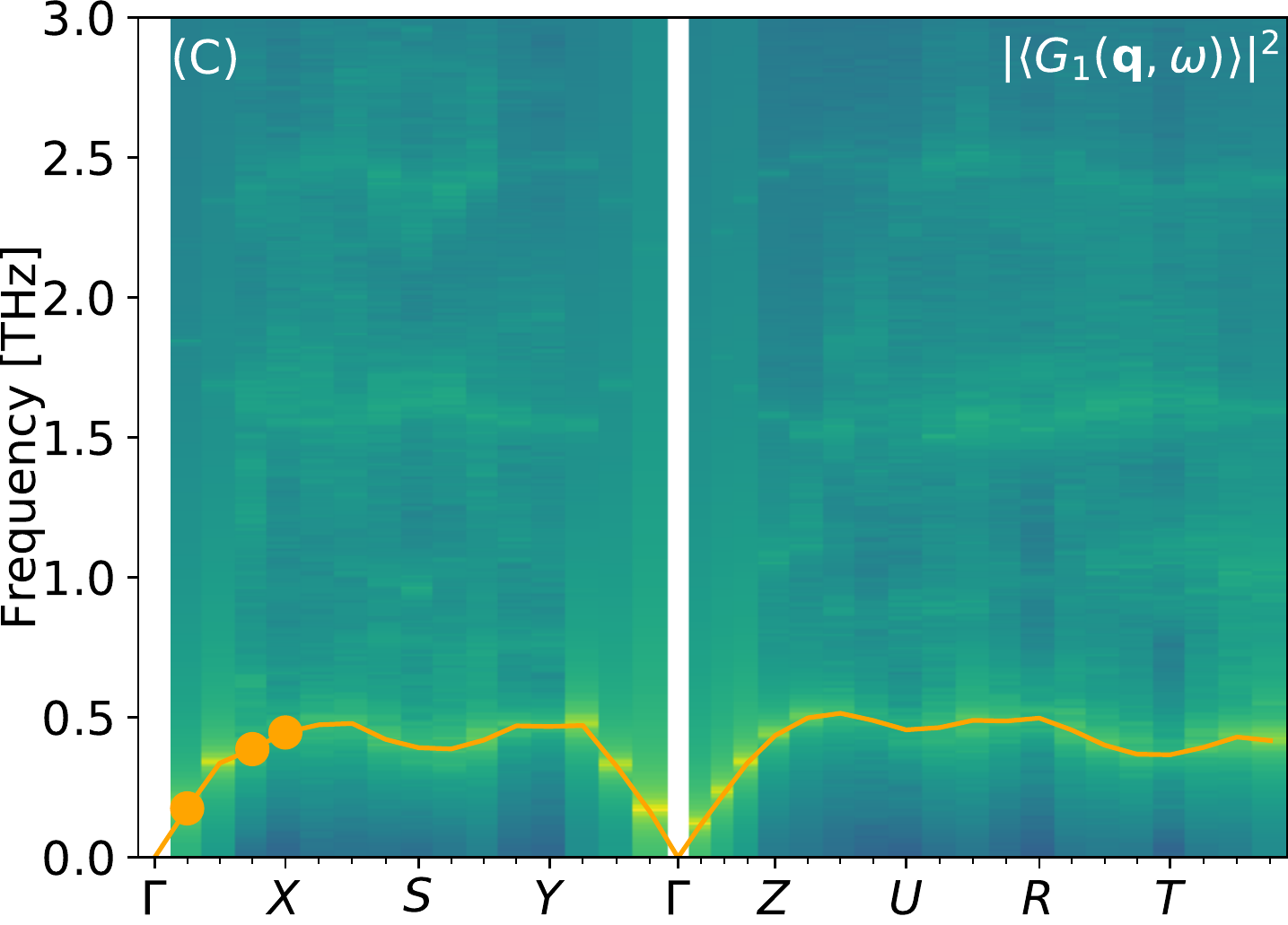}
	\includegraphics[width=1.0\columnwidth]{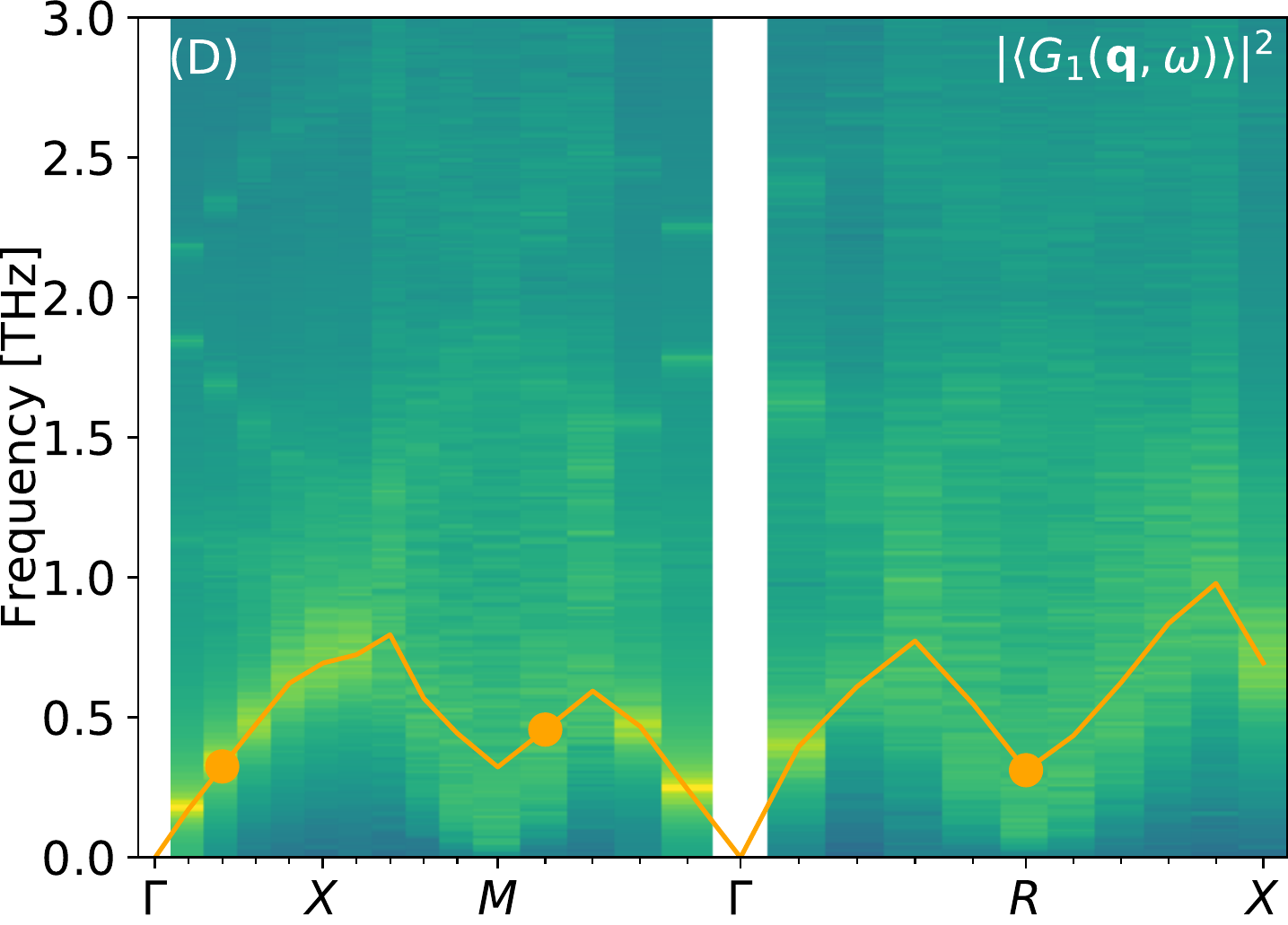}
	\caption{PVACFs of three selected $\mathbf{q}$-points projected on the
	eigenvectors of the first acoustic mode ($\alpha=1$)
	for (A) the orthorhombic phase at $150$~K and (B) cubic phase at 400~K.
	The solid lines show $|\langle G_{1}(\mathbf{q},\omega)\rangle|^{2}$
	and the dotted-dashed orange lines shows Lorentzian fitting.
	$|\langle G_{1}(\mathbf{q},\omega)\rangle|^{2}$ along the full
	$\mathbf{q}$-point path obtained for (C) 150~K and (D) 400~K. Orange
	circles denote renormalised frequencies $\tilde{\omega}_1(\mathbf{q})$
	corresponding to the peaks in (A,B) (from left to right).}
        \label{DynStrucFig10}
\end{figure*}

\subsection{The non-Lorentzian peak problem}\label{PPBPP}
When fitting the spectral densities obtained by Eq~(\ref{ProjectedPhonon4})
a problem with the peak shape appears, which indicates the strong
anharmonicity of the interaction potential.
This problem is visualised in Figure~\ref{DynStrucFig10} for both the
orthorhombic and cubic phase. Fig.~\ref{DynStrucFig10}(A) shows three peaks
corresponding to the PVACFs projected on eigenvectors of the first acoustic branch
($\alpha=1$) at three selected $\mathbf{q}$-points in the 150~K orthorhombic phase.
The solid lines are the $ \langle G_{1}(\mathbf{q},\omega)\rangle$
PVACFs power spectra and the dotted lines are the Lorentzian fits. The spectral
density of only $\alpha=1$ along the full high 
symmetry $\mathbf{q}$-path is shown in Fig.~\ref{DynStrucFig10}(C).
The orange circles in the plot are the
renormalised frequencies $\tilde{\omega}_1(\mathbf{q})$ corresponding to the
peaks shown in Fig.~\ref{DynStrucFig10}(A). For the
$\mathbf{q}=\left[\frac{1}{8}\,\,0\,\,0\right]$ peak a Lorentzian with a well-defined
eigenfrequency $\tilde{\omega}_{1}$ and linewidth $\Gamma_{1}$ can be fitted.
For the $\mathbf{q}=\left [\frac{3}{8}\ 0\ 0\right]$ peak resonances appear
as side peaks at higher frequencies in the spectrum. At $\mathbf{q}=\left [\frac{1}{2}\,\,0\,\,0\right]$
the power spectrum splits into two maxima emphasized by the inset of Fig~\ref{DynStrucFig10}(A).
The fitting results in a Lorentzian engulfing both of the peaks.
A well-defined weakly-interacting phonon quasi-particle should not show double peak
signals\cite{Zhang:prb2014,Sun:prb2014,Allen:prb2010,Ladd:prb1986,Lu:prl2017}.

Also at 400~K in the cubic phase, the peak problem appears as shown in
Figures~\ref{DynStrucFig10}(B,D). As before, only the acoustic branch $\alpha=1$ is shown in
Figures~\ref{DynStrucFig10}(D). The $[0\ \frac{2}{5}\ 0]$ peak shows a well-defined phonon
signal and can be fitted with a Lorentzian line shape. This is not the case for
the peaks related to the slightly-off-M point
$\mathbf{q}=\left[\frac{2}{5}\,\,\frac{2}{5}\,\,0\right]$ and
R-point $\mathbf{q}=\left[\frac{1}{2}\,\,\frac{1}{2}\,\,\frac{1}{2}\right]$. According to 
Ref.~\cite{Reissland:Book1973} a well-defined phonon has
to satisfy the following inequality,
\begin{equation}
	\tilde{\omega}_{\alpha}(\mathbf{q})\tau_{\alpha}(\mathbf{q}) > 1,
  \label{PeakCriterion}
\end{equation}
\noindent
where $\tilde{\omega}_{\alpha}(\mathbf{q})$ is the renormalized frequency and $\tau_{\alpha}(\mathbf{q})$
is the phonon lifetime. The phonons of the first acoustic branch with wave vectors
$\mathbf{q}=\left [\frac{2}{5}\ \frac{2}{5}\ 0\right ]$ and 
$\mathbf{q}=\left [\frac{1}{2}\ \frac{1}{2}\ \frac{1}{2}\right]$
do not satisfy this criterion and therefore don't form a well-defined state.
Additionally, the form of the peaks deviates from the Lorentzian
shape and would be better described by a skewed Gaussian distribution.
Therefore, also in this type of peak, the PVACF approach with harmonic
eigenvectors can not be applied in the strict sense of inequality~(\ref{PeakCriterion}).

Summarizing, phonons in CsPbBr$_3$ show many resonances with other phonon branches. 
Already in the spectral densities related to the first acoustic mode such resonances appear, as 
shown in Fig.~\ref{DynStrucFig10}. They show structure or high intensities not only
for the mean frequency, but also for a broad part of the frequency spectrum. The
further away from $\Gamma$ and the closer to the Brillouin zone boundary, the phonon quasiparticle in terms
of the PVACF picture becomes hardly applicable.

\begin{figure}
        \centering 
        \includegraphics[width=1.0\columnwidth]{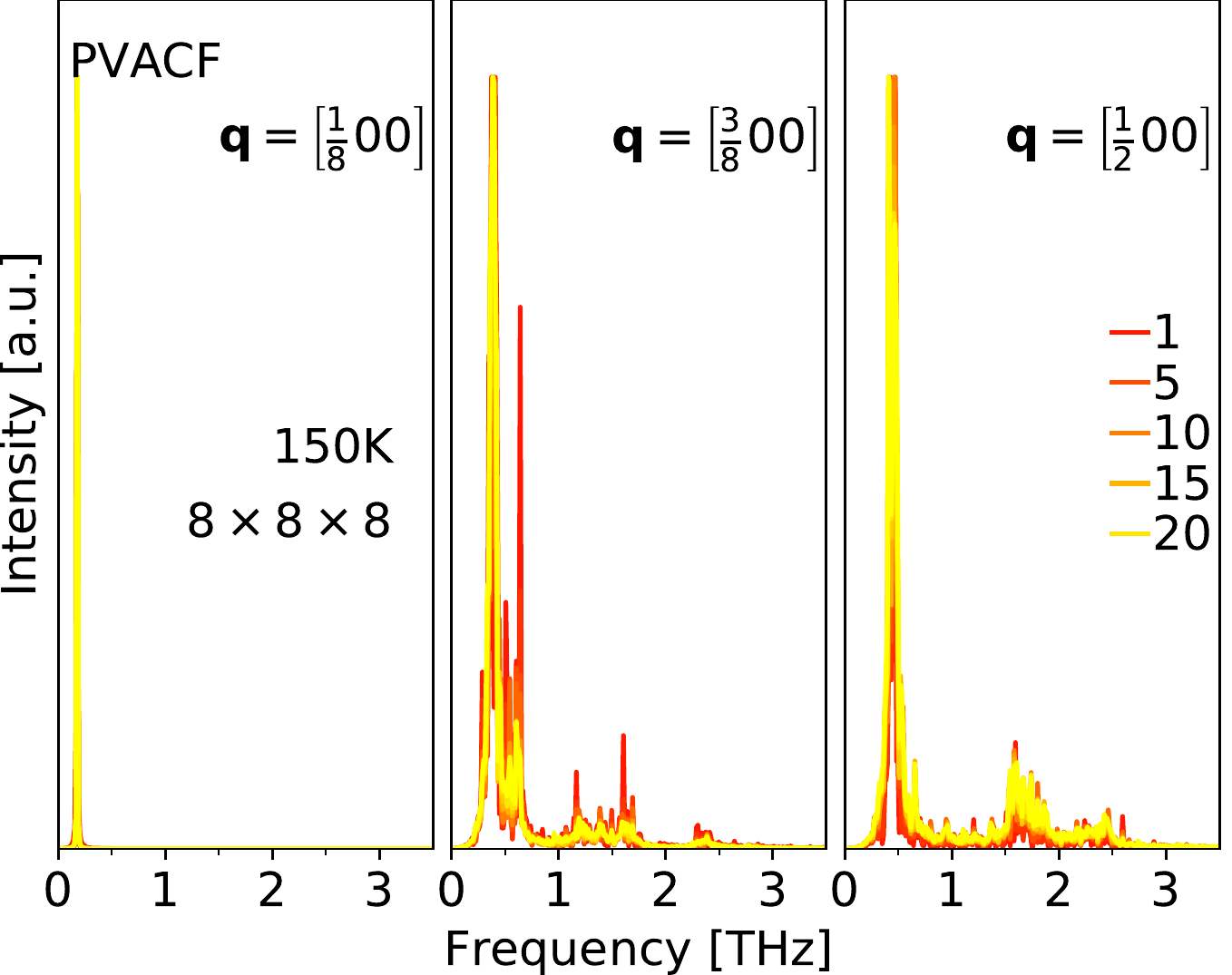}
        \includegraphics[width=1.0\columnwidth]{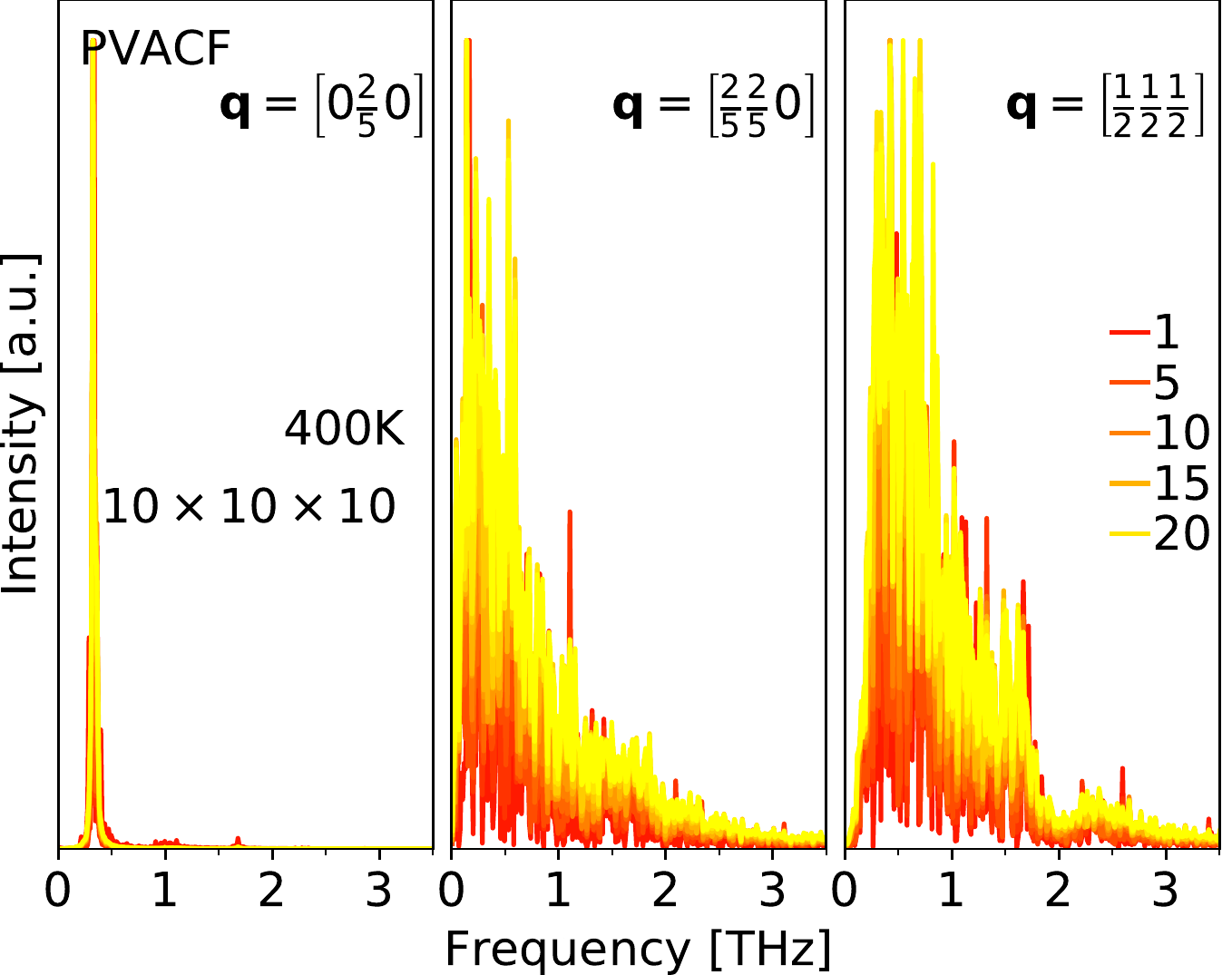}
        \caption{Convergence of the PVACF$_{\alpha=1}$ power spectrum.
	         The linear color scale of the curves depicts the number of trajectories,
		 used for the average. Red is a single trajectory and yellow
		 is an average over all 20 trajectories. Convergence of the signals for the orthorhombic 
		 ($8\times8\times8$) structure (\textit{top}), and for the cubic ($10\times10\times10$) phase (\textit{bottom}).
	         }
        \label{Convergencelarge}
\end{figure}
\begin{figure}[!b] 
        \centering 
        \includegraphics[width=1.0\columnwidth]{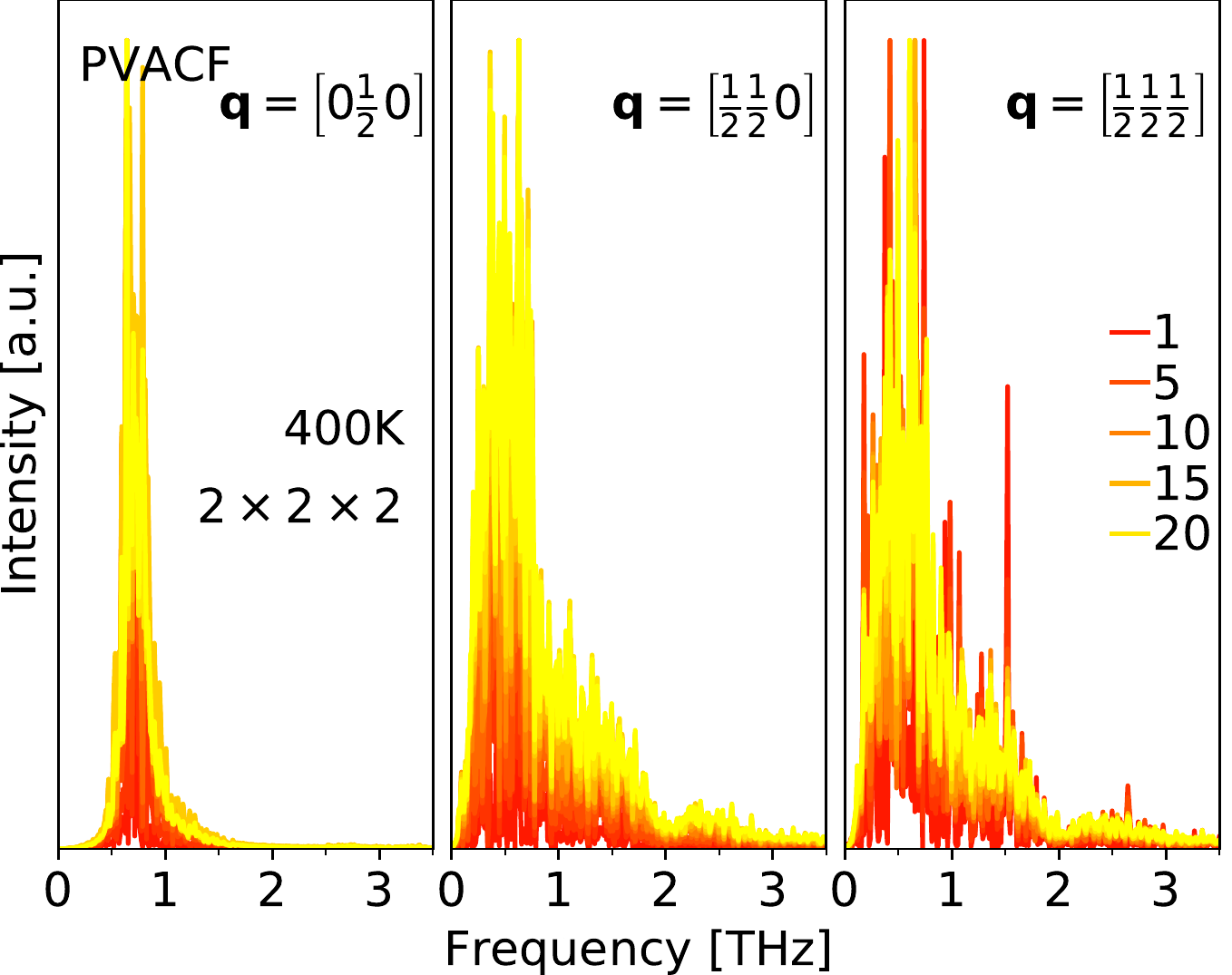}
        \caption{Convergence of the PVACF$_{\alpha=1}$ power spectrum in a cubic $2\times2\times2$ simulation
        at 400K. Color scheme as in Fig.~\ref{Convergencelarge}.}
        \label{Convergencesmall}
\end{figure}


\subsection{Convergence of the correlation functions}
\label{ConvergenceSection}
Before we start analysing the rest of the phonon spectrum (i.e. $\alpha\ge{}1$), we want to 
establish that our spectra are converged with respect to simulation time.
In Figure~\ref{Convergencelarge} the simulation time convergence of
the PVACF is shown for for $\alpha=1$ and the same $\mathbf{q}$-points as considered in 
Fig.~\ref{DynStrucFig10}. We are restricting our convergence analysis on the PVACF
under the assumption that if the individual phonon branches are converged so has to
be the $\mathbf{q}$-VACF which is a sum over the branch index $\alpha$.
The orthorhombic simulations are shown in the 
upper panel and the cubic results are depicted in the lower panel. The color of the lines 
denote the number of independent (200~ps long) trajectories over which the average was computed. 
Each starting structure and the related velocities of a trajectory were taken from a well-equilibrated $NVT$ ensemble. 
From red to yellow the curves show averages taken over 2, 4 to 20 $NVE$ trajectories.
The power spectra converge within the available amount of data of the PVACF. 
This can be seen because the double peak behaviour for the orthorhombic structure and the broad
skewed distributions for the cubic structures remain in the PVACF, but are 
getting smoother.
Notice the resonance peak intensities in Fig~\ref{Convergencelarge} at 
$\mathbf{q}=[\frac{3}{8}00]$ diminish during convergence.
This indicates that a large amount of data is needed to obtain correct peak intensities.
We also note that in the PVACF, peaks forming well-defined phonons converge faster
than the more deformed peaks which experience stronger phonon-phonon interactions.

To test the influence of our large simulation box, similar simulations were done for a small
$2\times2\times2$ structure at 400~K covering only $\frac{1}{125}$ of the volume of
the large box. This is the typical dimension of the simulation box used in first-principles based
MD simulations~\cite{Guo:acsel:2017,Hellmann:prl2020,Chai:prb2003,Zhang:prb2017,Zhang:prb2014,Sun:prb2014}.
The convergence results for the high symmetry $\mathbf{q}$-points are shown in Figure~\ref{Convergencesmall}.
Also for this system size the spectra converge within the available amount of data.
A comparison between the PVACF at 400~K in the $2\times2\times2$ (Fig.~\ref{Convergencesmall})
and the $10\times10\times10$ (Figure~\ref{Convergencelarge}) supercell at the 
$[\frac{1}{2}\frac{1}{2}\frac{1}{2}]$
$\mathbf{q}$-point shows equivalent signals for this $\mathbf{q}$-point. Only the relative noise in
the case of the $2\times2\times2$-cell is slightly larger. This is a direct result
of the smaller amount of samples when simulating a $2\times2\times2$-cell for the same amount of time.

\begin{figure}[!t] 
        \centering 
	\includegraphics[width=1.0\columnwidth]{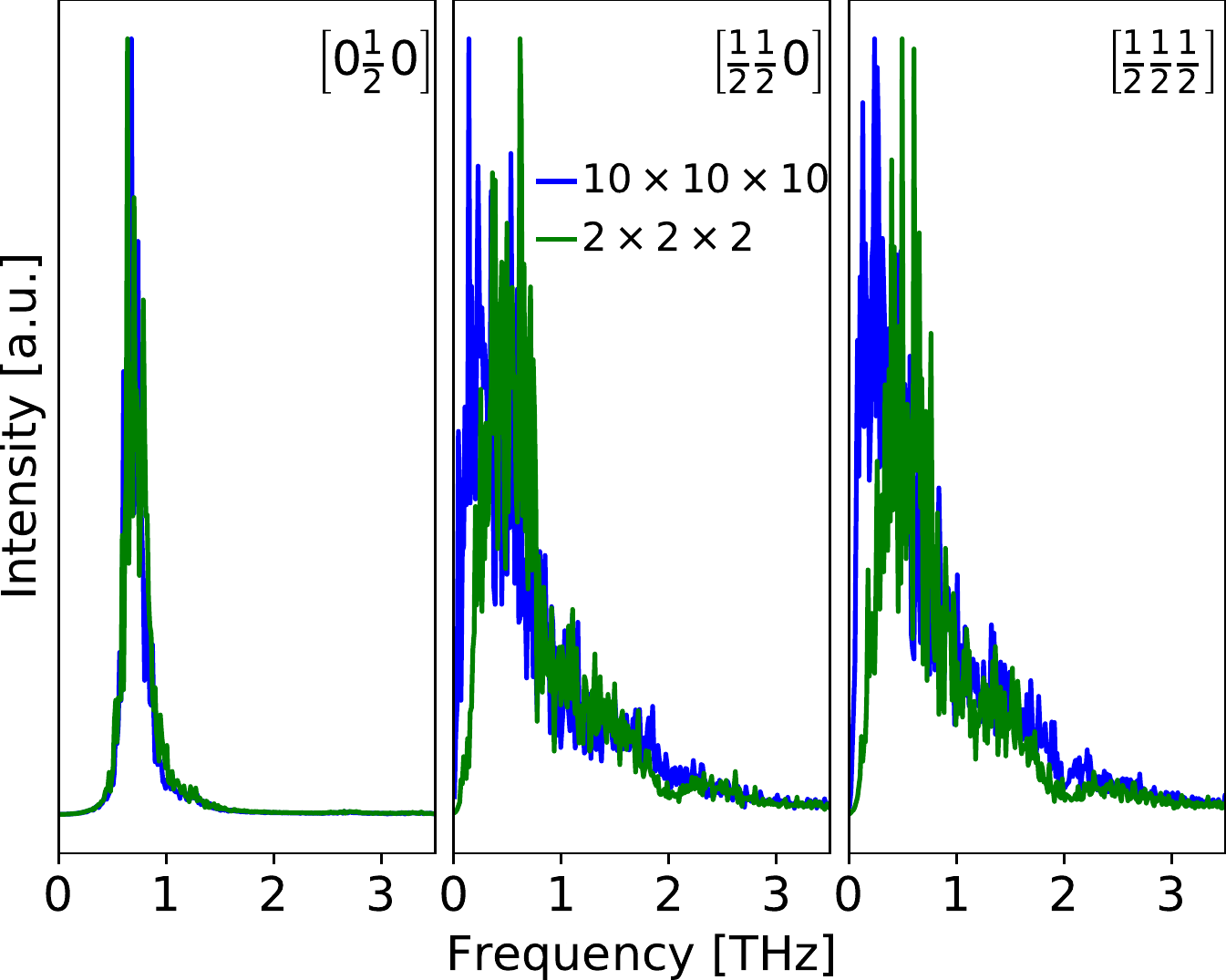}
	\caption{Comparison of the PVACF (${\alpha=1}$) power spectrum for a $10\times10\times10$ (blue) 
	and a $2\times2\times2$ (green) supercell in the 400~K cubic phase.} 
        \label{PVACFSizeAnalysisPlot}
\end{figure}

In Fig.~\ref{PVACFSizeAnalysisPlot} a comparison between the power spectra of the
first phonon branch in the $2\times2\times2$ and $10\times10\times10$ cubic phase is shown. 
Both boxes exhibit a narrow multi-peak signal $\mathbf{q}=[0\frac{1}{2}0]$, showing agreement
between the box sizes for this $\mathbf{q}$-point. For
$\mathbf{q}=[\frac{1}{2}\frac{1}{2}0]$ and $\mathbf{q}=[\frac{1}{2}\frac{1}{2}\frac{1}{2}]$,
differences between the $2\times2\times2$ and the $10\times10\times10$ are visible.
For both $\mathbf{q}$-points the peaks are narrower in the $10\times10\times10$  
box. For $\mathbf{q}=[\frac{1}{2}\frac{1}{2}0]$ the fitted line widths ($\Gamma_{\alpha}(\mathbf{q}))$
are $\sim0.40$~THz in the large box and $\sim0.46$~THz in the small box. For 
$\mathbf{q}=[\frac{1}{2}\frac{1}{2}\frac{1}{2}]$ a similar behaviour is observed,
where the $10\times10\times10$-box gives a linewidth of $\sim0.39$~THz and the small box
$\sim0.49$~THz. Most notably is the shift of the peak maximum to smaller frequencies in the larger box.
For $\mathbf{q}=[\frac{1}{2}\frac{1}{2}0]$ the renormalized frequency
in the $10\times10\times10$ box is $\sim0.32$~THz and in the $2\times2\times2$
box it is $\sim0.55$~THz, resulting in a frequency shift of $\sim0.21$~THz.
For $\mathbf{q}=[\frac{1}{2}\frac{1}{2}\frac{1}{2}]$ the renormalized frequency in
the larger box is $\sim0.31$~THz and $\sim0.55$~THz in the small box, what gives
a frequency shift of $\sim0.24$~THz similar to the previous $\mathbf{q}$-point. 
For the remaining acoustic and optical states we observe that if the powerspectrum 
of $G_{\alpha}(\mathbf{q})$ exhibits a narrow peak then the $10\times10\times10$ and 
the $2\times2\times2$ box give equivalent results as in the case of $\mathbf{q}=[0\frac{1}{2}0]$.
But for states that experience a higher degree of anharmonicity broader
peaks and higher frequencies are obtained in the $2\times2\times2$ box
compared to the $10\times10\times10$ box.

\begin{figure*}[!t] 
        \centering 
        \includegraphics[width=1.504\columnwidth]{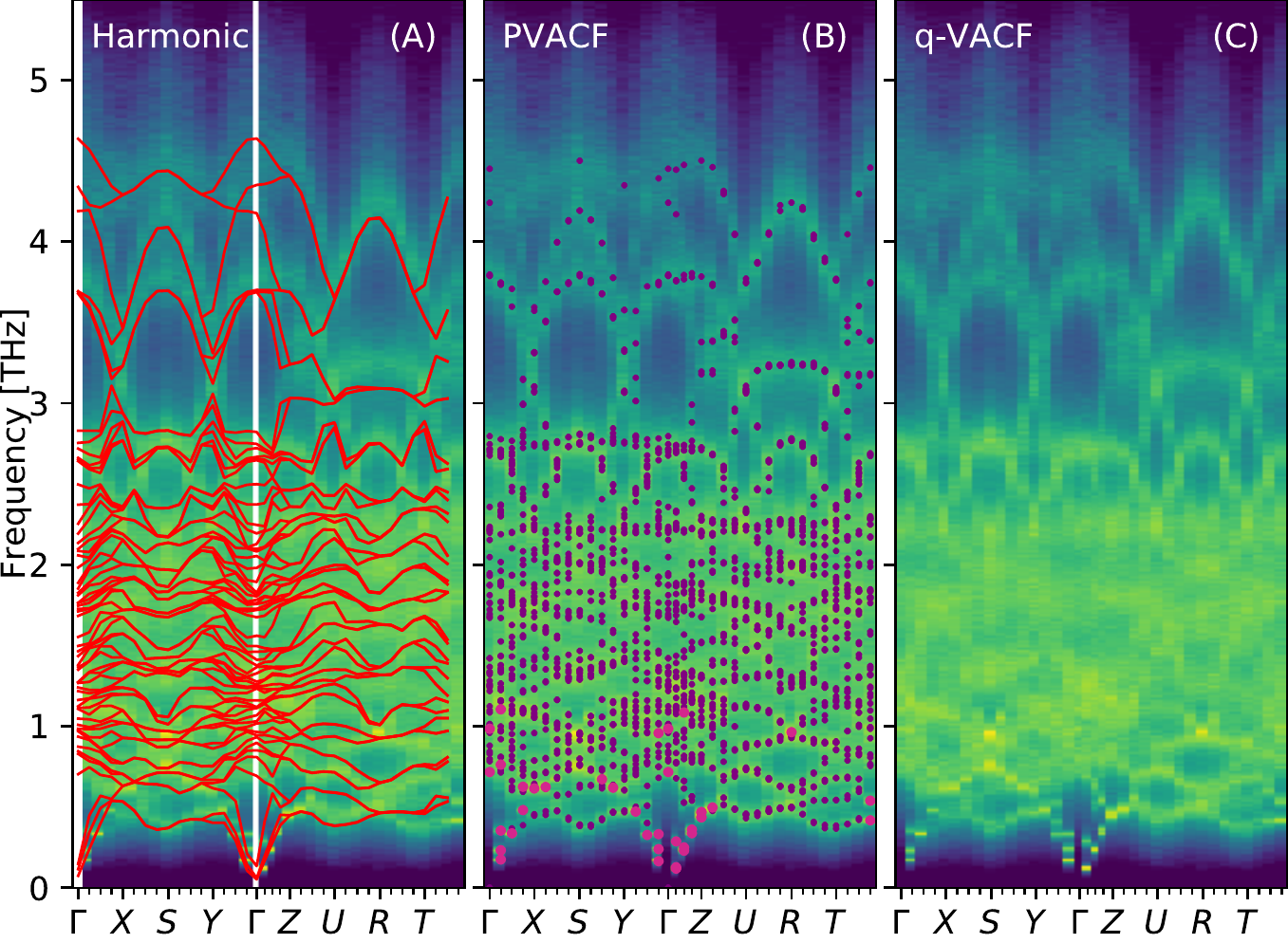}
        \caption{Phonon spectrum of orthorhombic CsPbBr$_3$ at 150~K. (a)
	The harmonic approximation is shown on top by the red lines. (b) the renormalized frequencies obeying 
	inequality~(\ref{PeakCriterion}) and not showing multi-peaks are shown in dark-purple,
	renormalized frequencies obtained by ignoring multi-peaks and ignoring inequality~(\ref{PeakCriterion})
	are shown by dark-purple dots. The background in (a) and (b) is the q-VACF as fully shown in (c).} 
        \label{DynStrucFig5}
\end{figure*}
\begin{figure*}[!t] 
        \centering 
        \includegraphics[width=1.504\columnwidth]{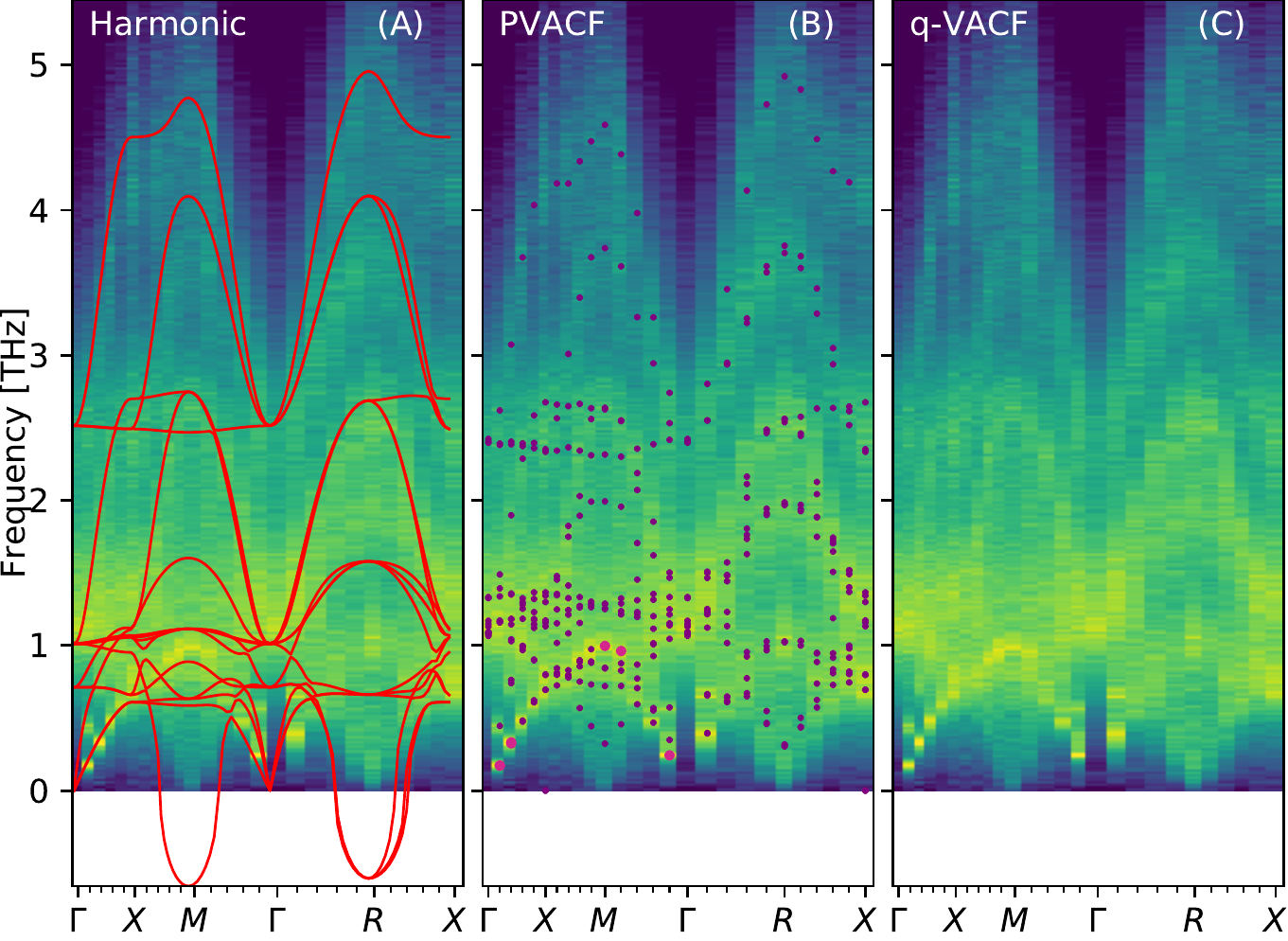}
        \caption{Phonon spectrum of orthorhombic CsPbBr$_3$ at 400~K. (a)
	The harmonic approximation is shown on top by the red lines. (b) the renormalized frequencies obeying 
	inequality~(\ref{PeakCriterion}) and not showing multi-peaks are shown in dark-purple,
	renormalized frequencies obtained by ignoring multi-peaks and ignoring inequality~(\ref{PeakCriterion})
	are shown by dark-purple dots. The background in (a) and (b) is the q-VACF as fully shown in (c).} 
	\label{DynStrucFig7}
\end{figure*} 

\begin{table}[!b] 
	\centering 
	\caption{Renormalized eigenfrequencies $\tilde{\omega}_{\alpha}(\mathbf{q})$ 
	         and harmonic phonon eigenvectors at high symmetry
        points corresponding to modes with imaginary eigenfrequency in harmonic 
	approximation.}
        \begin{tabular}{ | c | c | c | c | c |c| } 
		\hline 
		Mode  & & M$_1$  &   R$_1$ &   R$_2$ &  R$_3$ \\ 
                \hline 
		$\tilde{\omega}$ [THz] & & 0.09 & 0.27 & 0.22 & 0.27 \\ 
                \hline
                \multirow{3}{*}{Pb} &  x & 0.00 & 0.00 & 0.00 & 0.00 \\ 
		                    &  y & 0.00 & 0.00 & 0.00 & 0.00 \\ 
				    &  z & 0.00 & 0.00 & 0.00 & 0.00 \\ 
                \hline 
                \multirow{3}{*}{$\text{Br}_{x}$}
		          &  x & 0.00 & 0.00 & 0.00 & 0.00 \\ 
			  &  y & 0.70 & 0.47 & 0.00 & 0.52 \\
                          &  z & 0.00 &-0.50 & 0.10 & 0.48 \\ 
                \hline
                \multirow{3}{*}{$\text{Br}_{y}$}
		          &  x & -0.70 & -0.47 & -0.11 & -0.52 \\
			  &  y & 0.00 & 0.00 & 0.00 & 0.00 \\
                          &  z & 0.00 & -0.14 & 0.69 & 0.00\\ 
                \hline \multirow{3}{*}{$\text{Br}_{z}$}
		          &  x & 0.00 &0.51 & 0.11 & -0.48\\ 
			  &  y &  0.00 &0.15 &-0.69 & 0.00\\ 
			  &  z &  0.00 &0.00 & 0.00 & 0.00\\ 
                \hline \multirow{3}{*}{Cs}
                          &  x &  0.00 &0.00 & 0.00 & 0.00\\ 
                          &  y &  0.00 &0.00 & 0.00 & 0.00\\ 
			  &  z &  0.00 &0.00 & 0.00 & 0.00\\ 
                \hline 
        \end{tabular}
        \label{MRpointTable} 
\end{table}

\subsection{CsPbBr$_3$ PVACF and q-VACF power spectra analysis}
\label{DispersionSection}

The dispersion relation for the orthorhombic phase including all acoustic and optical modes 
computed with the harmonic approximation is shown in Figure~\ref{DynStrucFig5}(A) by 
the red lines.
The renormalized frequencies obtained from the PVACF are shown in Figure~\ref{DynStrucFig5}(B)
by the dark and light-purple circles. The dark-purple points
denote the PVACF power spectra for which the Lorentzian fitting was performed 
within the bounds of inequality~(\ref{PeakCriterion}) and without
experiencing problems with \textit{multi-peak} spectra. The light-purple points
were obtained when ignoring the restrictions imposed by the used method.
It is remarkable how few points in the bandstructure satisfy 
inequality~(\ref{PeakCriterion}) and show no resonance peak(s).
Nevertheless, a comparison by eye of the so obtained renormalized frequencies to the colored 
background of the q-VACF seem to qualitatively agree with the dispersion 
obtained by the harmonic approximation. However, the average root-mean-square 
(rms) error computed between harmonic frequencies $\omega_{\alpha}(\mathbf{q})$
and the renormalized frequencies $\tilde{\omega}_{\alpha}(\mathbf{q})$ at the same 
$(\mathbf{q},\alpha)$ shows a large difference. Its large value of 0.3~THz indicates that 
the used fitting procedure has troubles with the many multi-peak features in the 
spectrum. This was exemplified by the inset of Fig~\ref{DynStrucFig10}(A).

In the q-VACF we can see overlapping flat bands up to a frequency range of around 3~THz. 
Overall, the broad and flat character of the bands suggests low phonon lifetimes
and a high degree of anharmonicity, which is in agreement with previous studies
\cite{Simoncelli:natp2019,Langian-Atkins:natm2021,Guo:acsel:2017,Lee:pnas2017}.


The analysis of the cubic phase at $400$~K is shown
in Figure~\ref{DynStrucFig7}. The style of the figure is the same as that of Fig.~\ref{DynStrucFig5}.
The renormalized frequencies 
$\tilde{\omega}_{\alpha}(\mathbf{q})$ obtained from the PVACF are depicted in
Figure~\ref{DynStrucFig7}(B).
The harmonic approximation Fig~\ref{DynStrucFig7}(A) shows imaginary eigenfrequencies for 
a single acoustic band at $\mathbf{q}=M$ and for all three
acoustic modes at $\mathbf{q}=R$. The eigenvectors belonging
to the imaginary eigenfrequencies are tabulated in table~\ref{MRpointTable}.
These eigenmodes belong to polar rotational modes of the 
PbBr$_{6}$ octahedra around the enclosed lead atoms. Only Br atoms are participating in
those stabilizing modes. By renormalizing the frequencies in the PVACF approach we obtain with real valued
frequencies $\tilde{\omega}_{\alpha}(\mathbf{q})$. In this process the 
three-fold degeneracy of the acoustic mode at the R point is lifted.

As in the orthorhombic phase there are very few spectra that can be
properly fitted, in the sense that it obeys the limits of inequality~(\ref{PeakCriterion}).
Both the harmonic approximation and the PVACF show flat and close lying
bands between 0.5 and 1.5~THz. The rms error between the harmonic
approximation and the PVACF approach is 0.5~THz, i.e. slightly larger to the
orthorhombic phase, what is expected for higher temperatures. The MD approach 
shows phonon mode softening for the two highest lying optical
bands, compared to the harmonic approximation. 
The q-VACF is shown in Figure~\ref{DynStrucFig7}(C) and exhibits an overlapping band spectrum over
the whole frequency region with only little structure. This
indicates a high degree of anharmonicity in the underlying potential. 
\begin{figure}[!t] 
        \centering 
	\includegraphics[width=1.0\columnwidth]{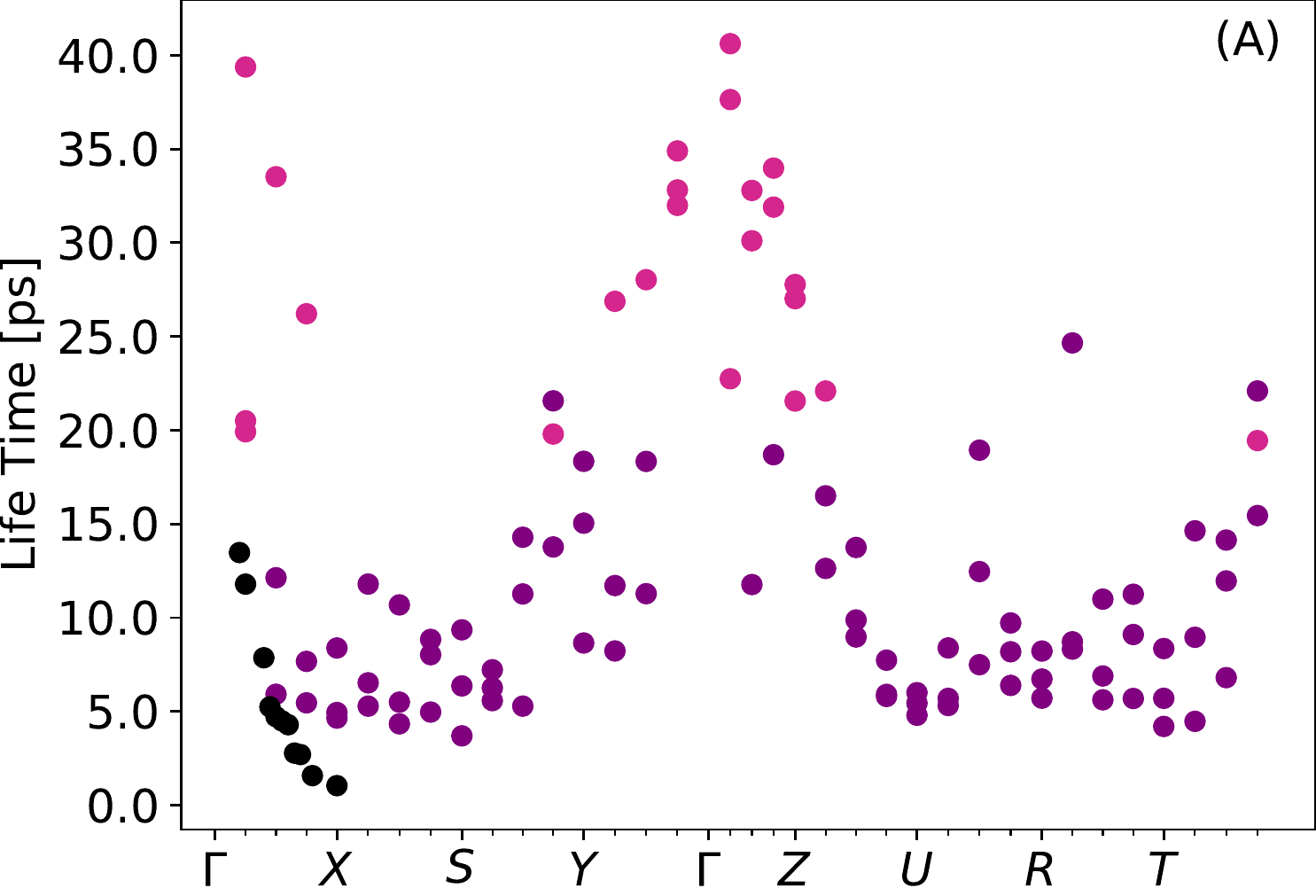}
	\includegraphics[width=1.0\columnwidth]{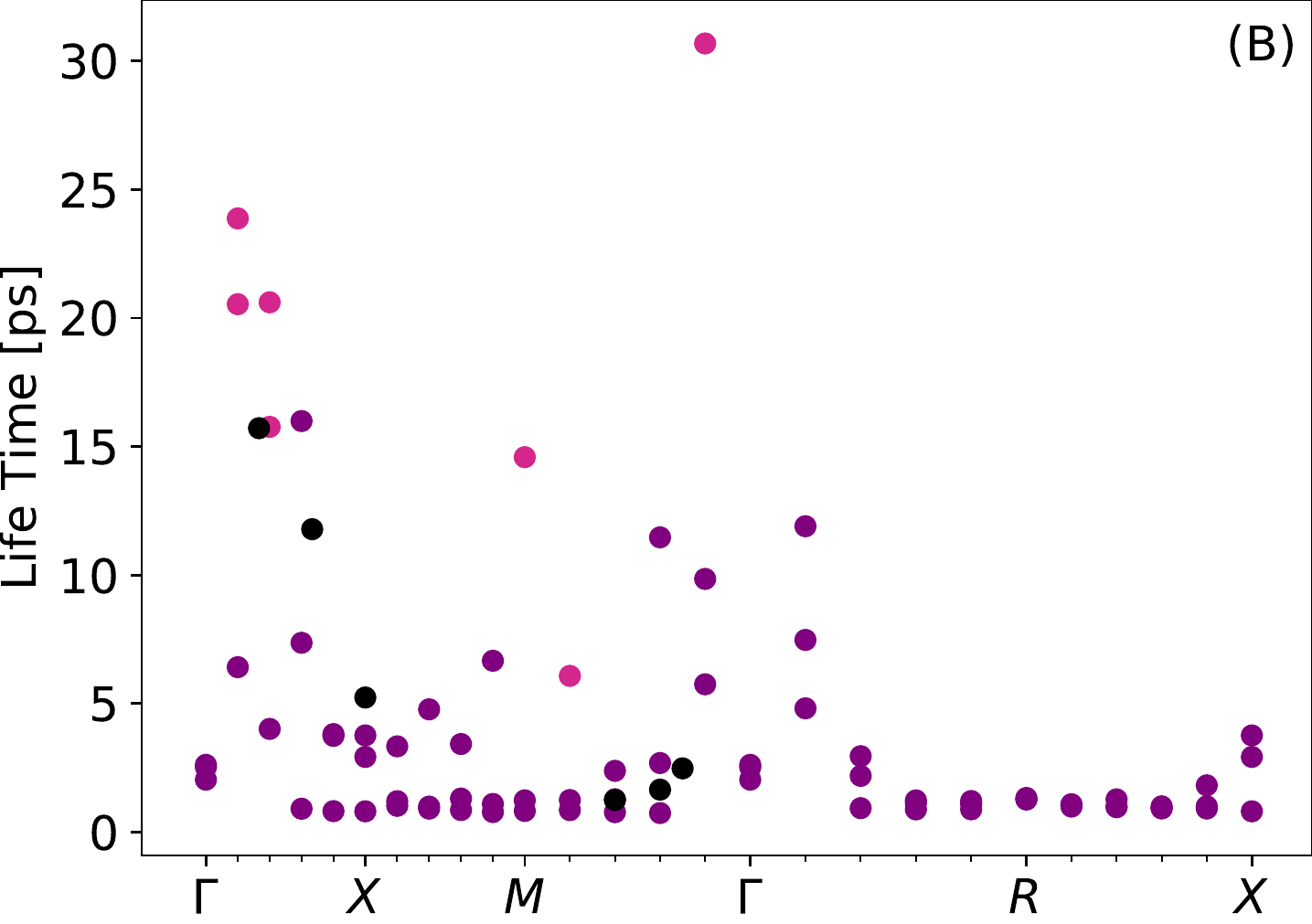}
	\caption{Acoustic phonon ($\alpha=1,2,3$) lifetimes 
		 for the (A) orthorhombic and
		 the (B) cubic system obtained from the PVACF approach.
                 The dark-purple circles depict the accurately fitted MD data.
		 The light-purple circles represent peaks that were fitted by
		 violating the proper phonon picture. The black circles show experimental
		 phonon lifetimes of Ref.~\cite{Songvilay:prm2019}.}
        \label{PhononLifeTimesPlot}
\end{figure} 
\noindent


\subsection{Acoustic phonon lifetimes a comparison to experiment}

The phonon lifetimes for the three acoustic modes were obtained by Eq.~(\ref{PhononLifeTimes}) 
from the fitted PVACF line widths $\Gamma_{\alpha}(\mathbf{q})$.
The results are shown in Figure~\ref{PhononLifeTimesPlot}. Meaningful
lifetimes in terms of inequality~(\ref{PeakCriterion})
could only be obtained around the 
$\Gamma$ point, as shown by the light-purple circles. 
The lifetimes indicated by the dark-purple circles were obtained by
ignoring inequality~(\ref{PeakCriterion}) and the resonance peaks. 
Figure~\ref{PhononLifeTimesPlot}(A) shows three points for
every $\mathbf{q}$-point corresponding to the lifetimes of the acoustic 
branches $\alpha=1,2,3$ in the orthorhombic phase. 
Additionally, experimental data for the acoustic modes measured by Songvilay~\textit{et.al.}\cite{Songvilay:prm2019} is shown
by the black circles. The experimental values and the phonon
lifetimes measured from MLFF MD are of the
same order of magnitude, and show the same trend when moving away from $\Gamma$.
Interestingly, the experiment was only able to resolve
phonon lifetimes in roughly the regions of the $\mathbf{q}$-space where our approach was
able to predict lifetimes in agreement with inequality~(\ref{PeakCriterion}). 
Figure~\ref{PhononLifeTimesPlot}(B) shows that the cubic phase has on average 
shorter acoustic phonon lifetimes. Also here, qualitative agreement with 
experiment\cite{Songvilay:prm2019} is found for the acoustic modes along the $\Gamma{\rm -X}$ and $\Gamma{\rm -M}$ paths.

For both the orthorhombic and the cubic simulations meaningful results can only be
obtained for acoustic phonons with low momentum, ie. close to the $\Gamma$ point.
The average acoustic phonon lifetimes are significantly higher in
the orthorhombic compared to the cubic phase. 
This was to be expected since the atoms experience a lower degree of anharmonicity, 
as their displacements from equilibrium are smaller at lower temperatures.

\subsection{Atom type resolved power spectra}
\label{DOSSection}

\begin{figure*}[!t] 
    \includegraphics[width=1.0\columnwidth]{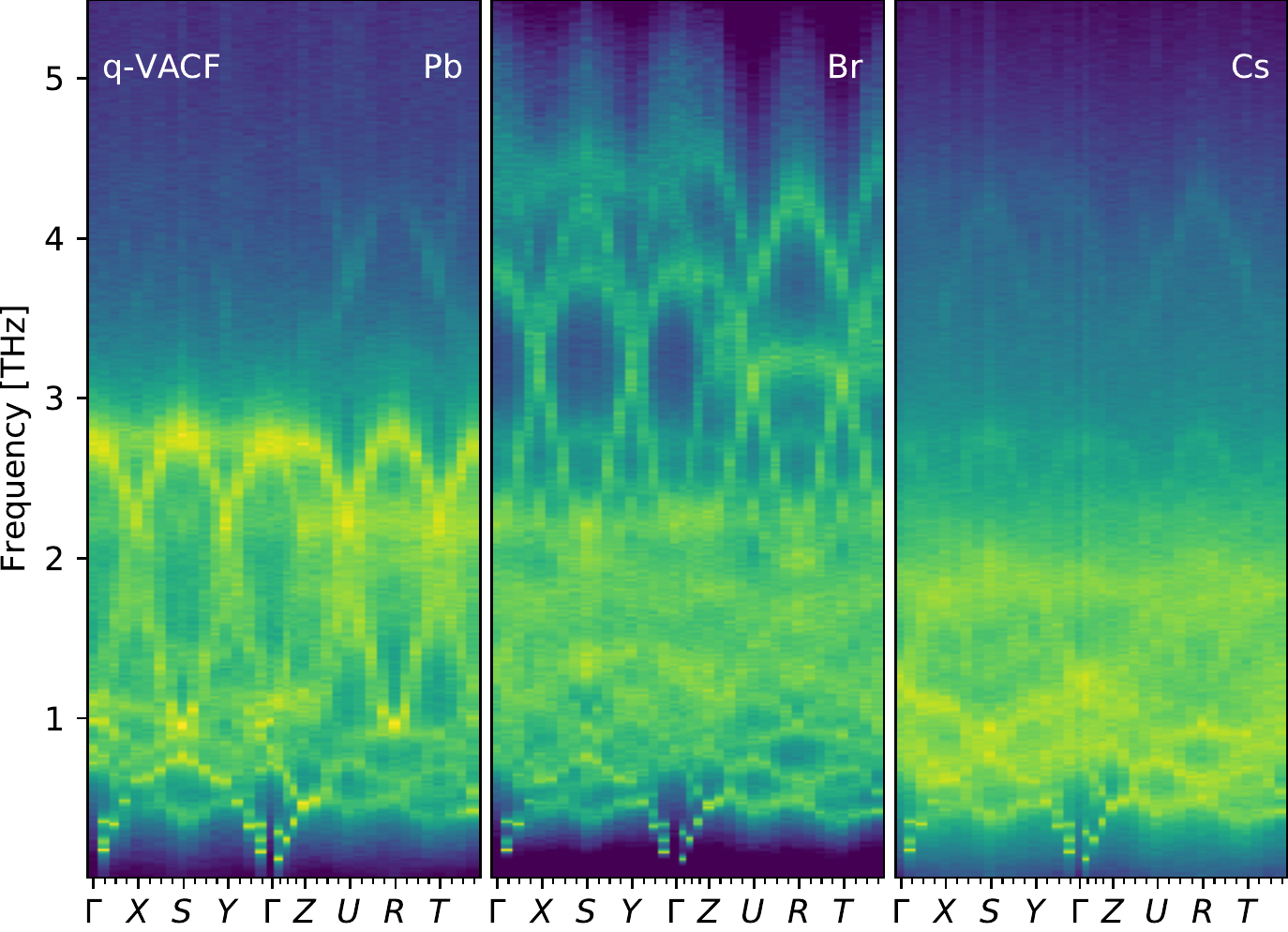}
    \includegraphics[width=1.0\columnwidth]{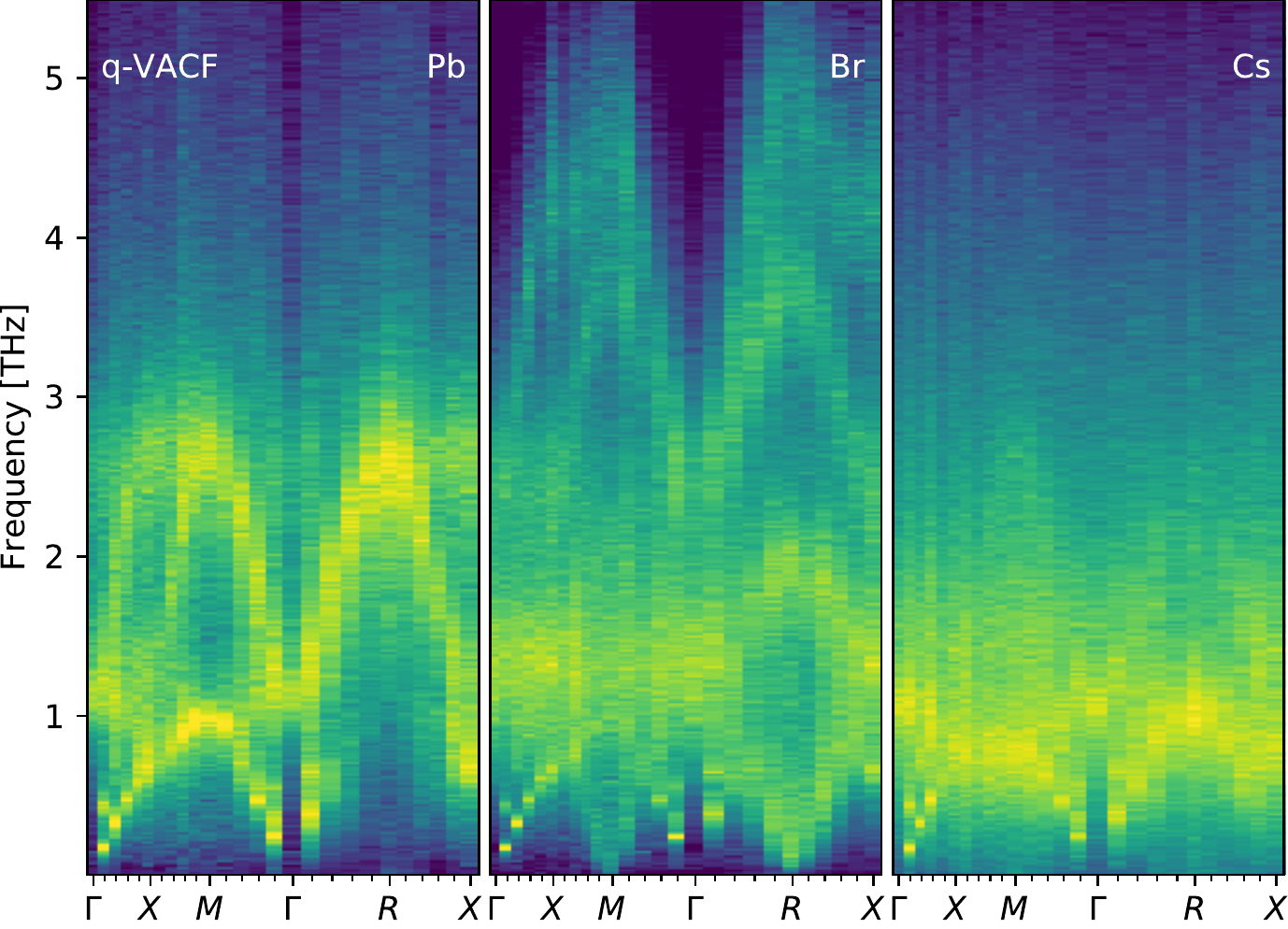}
    \caption{Atom resolved q-VACF for the (\textit{left}) orthorhombic phase at 150~K and (\textit{right}) cubic phase at 400~K.}
    \label{ArDsfVacf}
\end{figure*}

As Figs.\ref{DynStrucFig10},~\ref{DynStrucFig5}\&{}\ref{DynStrucFig7} have shown, the q-VACF and PVACF,
show many highly anharmonic modes overlapping in the $[0.5,3]$~THz range. This makes the 
spectra difficult to disentangle. Therefore, a $\mathbf{q}$ and atom
resolved form of q-VACF is shown in Figure~\ref{ArDsfVacf}. For both the cubic
and the orthorhombic phase, the atom resolved q-VACF displays sharper features
compared to the full q-VACF. A most remarkable feature is the broad flat band around $\sim$0.8~THz
for the Cs$^{+}$ cations extending over the whole Brillouin zone. This band is visible in both
phases. In particular the q-VACF approach applied to the
cubic phase visualizes this 'rattling band'. This band is
nearly dispersionless around the rattling frequency (0.8~THz) obtained in Section~\ref{sec:results:rattle}.

\begin{figure}[t!] 
        \centering 
	\includegraphics[width=1.0\columnwidth]{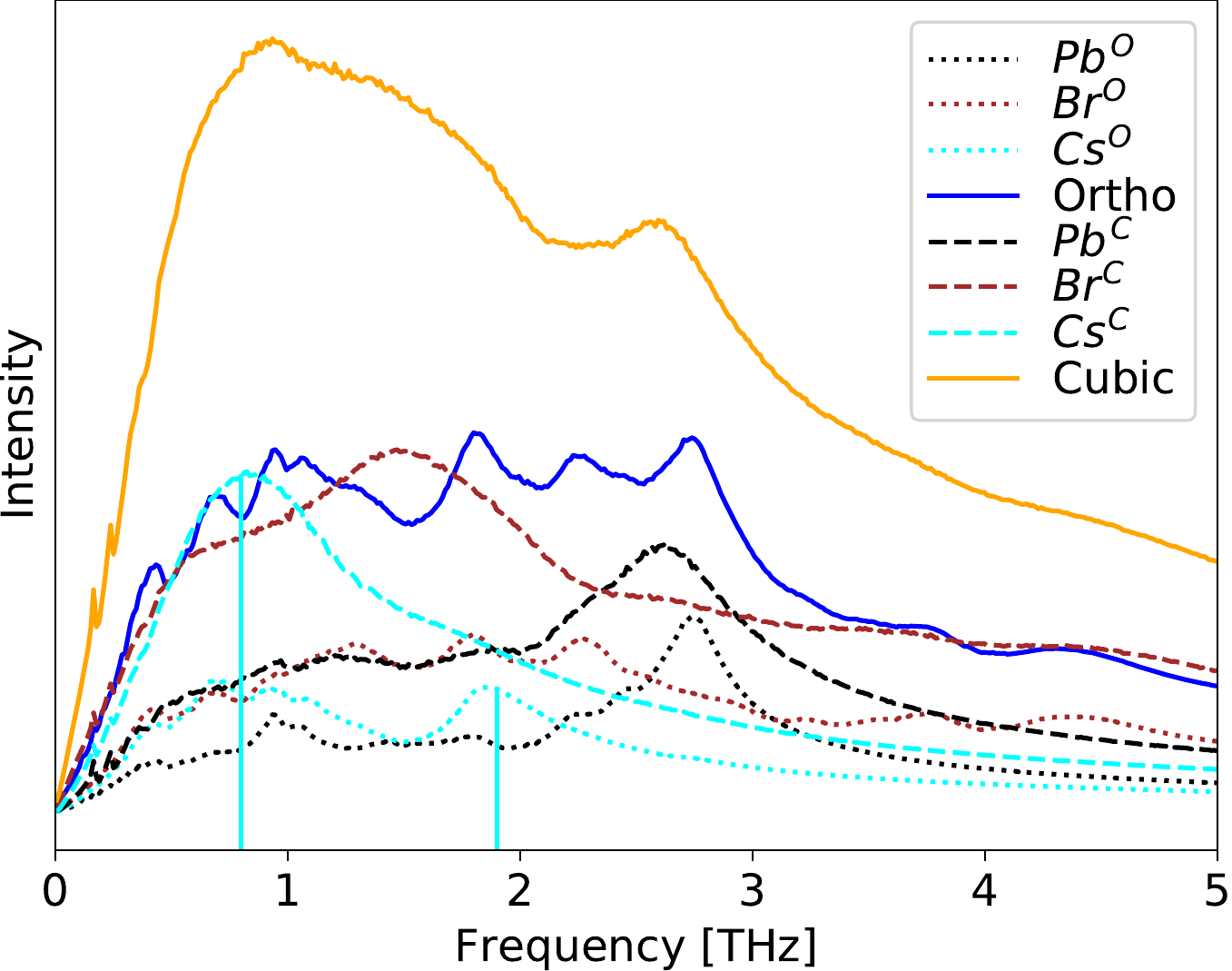}
        \caption{Phonon density of states obtained from unprojected 
	         velocity autocorrelation functions.
		 The letter O in the legend denotes orthorhombic and
		 C denotes cubic simulations.
		 The vertical light-blue
		 lines are the Cs rattling frequencies.
		 }	
        \label{PhononDOSPlot}
\end{figure}

The phonon density of states and its atom decomposed form are computed
and shown in Figure~\ref{PhononDOSPlot} for the orthorhombic and the cubic
phase. We now clearly see that the low frequency peak
of the Cs$^{+}$ cations coincides with the rattling frequency. 
This Cs peak makes an important contribution
to the total phonon DOS of the material.
In the orthorhombic simulation there is 
another Cs peak visible at 1.9~THz, which is absent in
the cubic simulations. This peak can not be assigned
to the random thermal vibrations determined in Section~\ref{sec:results:rattle}. 
This Cs peak is related to optical modes, and also shows the signatures of a dispersionless 
rattling band.

Furthermore, we can see the effect of the orthorhombic and cubic phase on the lead and bromide atoms.
The DOS of the lead atoms show a peak at 2.75~THz, which is also visible 
in the total DOS. This peak is present in both phases, and
experiences a red-shift of $\sim0.1$~THz when going from orthorhombic
to the cubic phase. In the bromide DOS of the orthorhombic
phase multiple peaks are observed below 3~THz, which are smeared out in the cubic phase.

\subsection{Disentangling phonon resonances in the acoustic branches}

\begin{figure}[!h] 
        \centering 
	\includegraphics[width=1.0\columnwidth]{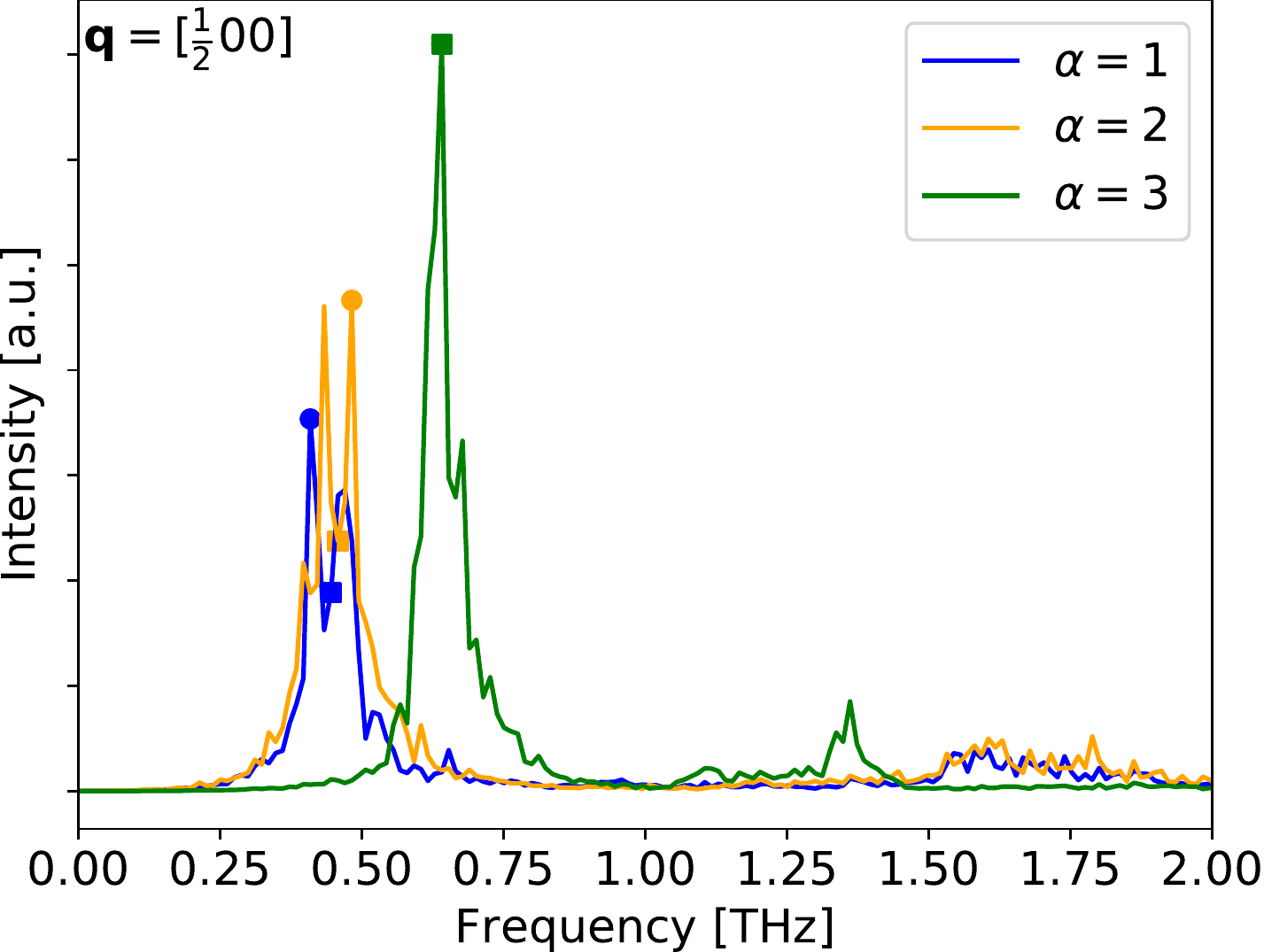}
	\caption{Power spectra of acoustic phonons measured in the orthorhombic phase
		 at $\mathbf{q}=[\frac{1}{2}00]$. Circles show the maximum peak. The square symbols
		 show the fitted peak maxima, i.e. the renormalized 
		 frequencies $\tilde{\omega}_{1}(\mathbf{q})$} 
        \label{PhononInteraction}
\end{figure}

Before discussing the results and formulating our conclusions we will
analyse the \textit{multi-peak} spectra observed in
the orthorhombic structure in more detail. We consider
the three acoustic phonon branches at $\mathbf{q}=[\frac{1}{2}00]$, 
shown in Figure~\ref{PhononInteraction} as representatives.
Phonon resonances can also occur between different $\mathbf{q}$-vectors,
but as a simplification we will restrict ourselves to a single $\mathbf{q}$-point.
All three acoustic signals show multi-peak behaviour. The positions marked with
squares denote the locations of the
renormalized eigenfrequencies. Beside the \textit{eigen-peaks} every signal shows a
number of \textit{resonant-peaks}. The first acoustic phonon branch ($\alpha=1$)
shows a \textit{multi-peak} at $\sim0.45$~THz. The second acoustic branch shows a very similar behaviour,
with \textit{multi-peaks} around the fitted frequency of $\sim0.47$~THz. The two close lying peaks
and the close lying eigen-frequencies of these two states indicate that they are very likely to resonate.
In terms of the first acoustic branch we could argue that the peak higher in intensity
is the phonon peak and the smaller one is the resonance with phonon branch 2. This argumentation
can not be used for the second branch $\alpha=2$ because in this case the peaks are equal in magnitude
and no unique choice what the main peak should be can be made. Moreover, both signals $\alpha=1$ and
$\alpha=2$ show small resonances roughly at the positions of the main peak of the third acoustic branch
($\alpha=3$). The third acoustic branch has its eigenfrequency located at $\sim0.6$~THz. It shows a resonant peak
($\sim0.7$~THz) that can not be assigned to any of the acoustic branches at the same $\mathbf{q}$-point. All three 
acoustic branches show resonances with higher lying optical modes at the same $\mathbf{q}$-point 
with frequencies above $1.2$~THz. This indicates same $\mathbf{q}$-point interaction of phonon states with 
frequencies lying more than $1$~THz appart.

\noindent
From this analysis it follows that a decoupling of the atomic motions into independent 
phonon eigenmodes is not possible for this material. The potential is too anharmonic and the 
spacing between the phonon bands in the orthorhombic phase is too small to avoid resonances. Therefore
it is not possible to decouple the system into independent modes. These phonon-phonon
resonances are not only visible in the acoustic branch but are occurring throughout the
whole phonon spectrum. Especially in regions of close lying bands and high crystal momentum
$|\mathbf{q}|$, the power spectra show more resonances between phonon modes
$\alpha$ and $\alpha'$.

\section{Discussion and Conclusion}\label{DisConc}
In this work it was shown that large-scale MD
simulations with accurate machine-learning potentials 
enable realistic simulations of the lattice dynamics
of highly anharmonic materials such as the CsPbBr$_{3}$ perovskite. The importance of phonon-phonon interactions
for CsPbBr$_{3}$ was reported in the
computational work of Simonicelli~\textit{et.al.}~\cite{Simoncelli:natp2019}. Here, we showed that large
supercell sizes are required in order to capture all phonon-phonon interactions 
properly. This was verified by our finite size analysis comparing the PVACF calculations
of a $2\times2\times2$ and a $10\times10\times10$ supercell. This showed that depending 
on the phonon state the anharmonicity the phonon is experiencing
can vary with the system size (Fig~\ref{PVACFSizeAnalysisPlot}).
The statistical convergence analysis showed that a large amount of 
MD samples are needed to properly converge the power spectra. The large-scale MD
could not have been carried out with first-principles based
MD in previous 
works~\cite{Guo:acsel:2017,Gehrmann:natc:19,Hellmann:prl2020,Chai:prb2003,Zhang:prb2017,Zhang:prb2014,Sun:prb2014}.
Larger simulations would be computationally too
expensive and only small supercells are tractable. The applied machine-learning approach\cite{Jinnouchi:prl19}
combined with the here developed analysis code ``\href{https://github.com/Jonathan271828/DSLEAP}{DSLEAP}''
opens up the possibility to study anharmonic 
lattice dynamics by large-scale MD simulations and taking into account
all degrees of anharmonicity. The extracted acoustic phonon lifetimes qualitatively agree with 
experimental results~\cite{Songvilay:prm2019}; they are of the same order of magnitude and show
similar $\mathbf{q}$-dependence. The phonon lifetimes are inversely 
proportional to their distance from $\Gamma$.

We conclude that 
it is not possible to decompose the lattice dynamics of CsPbBr$_{3}$ into a set of independent
oscillators. Inequality~\ref{PeakCriterion} only holds for peaks close to the $\Gamma$ point in the orthorhombic
and only for the acoustic modes close to the
$\Gamma$ point in the cubic phase. This shows that phonon-phonon interactions play a major role in this
system and cannot be ignored. Therefore we propose that the phonons in CsPbBr$_{3}$ have to be considered as
a phonon-liquid and not as a weakly interacting gas. This can be clearly recognized by the
strong resonances observed in the power spectra indicating coupled phonon states. 
We are using the term phonon-liquid as an analogy between phonons and particles in real space. 
If real space particles are very diluted, then their interactions can be 
neglected and they form a gaseous state. This
is analogous to phonons in a harmonic solid. The phonons are only very weakly interacting and
form a so called phonon gas\cite{Reissland:Book1973}. If the density of real space particles
is raised, the interactions become important and we are talking about a liquid.
The analogous behaviour in the phonon picture is a highly anharmonic crystal.
The anharmonicities introduce strong phonon-phonon interactions that can not be ignored and 
hence we are talking about a phonon-liquid. The importance of 
phonon-phonon interactions was also recognized in Ref.~\cite{Simoncelli:natp2019} and in
the neutron scattering study of Ref.~\cite{Langian-Atkins:natm2021}. Similar results
were obtained in a combined experimental and theoretical work
by Sharma \textit{et.al.} for the MAPbI$_3$ perovskite\cite{Sharma:prm2020}.
In this study we attempted to identify main and resonance peaks by comparing the acoustic phonon
branches for a selected $\mathbf{q}$-point (see Fig.~\ref{PhononInteraction}). The intensity
of these resonant peaks can be used as a measure for the importance of the phonon-phonon interaction. 
In some cases the resonant peaks are nearly as intense
as the main peak. These findings indicate a high degree of anharmonicity in the underlying potential. 

In the literature the ultra-low thermal conductivity is partly
attributed to 'rattling' motions of the Cs$^{+}$ cations\cite{Lee:pnas2017,Simoncelli:natp19}.
Rattling motions are also said to be responsible for low thermal 
conductivities in other materials as for example in sodium 
cobaltate~\cite{Vonshen:natm2013} or CuCrSe$_{2}$~\cite{Niedziela:natp2019}. The study of the Cs$^{+}$ cation correlation
function showed that rattling frequencies can be extracted from MD runs by 
the use of correlation functions of atomic displacements. A rattling frequency of $0.8$~THz
and a faster random motion of roughly $2.5$~THz, characterizes this process.
The atom decomposed $\mathbf{q}$-VACF show that rattling motions show only very weak dispersion. 
Therefore, we define rattling motions as a nearly dispersionless atomic oscillation
with a broad frequency spectrum, but with a well defined average frequency.


The analysis of the dispersion curves has shown that the CsPbBr$_3$ is dynamically
stabilized. Dynamic stabilization
is indicated by imaginary modes in the harmonic approximation that can 
be renormalized by the PVACF approach to give positive frequencies.
This finding is in agreement with the computational studies of
Refs.~\cite{Langian-Atkins:natm2021,Guo:acsel:2017,Gehrmann:natc:19}.
The modes responsible for the dynamic stabilization are the M and R points, and contain only
motions of the Br atoms. These modes form the characteristic rotation and tilting
pattern known to be important in many perovskite structures. For example, in the 
CaSiO$_{3}$~\cite{Sun:prb2014} or MAPbI$_{3}$~\cite{Whalley:prb2016,Sharma:prm2020} perovskites.

Summarizing, we have shown that the phonon properties of such highly anharmonic
materials as the CsPbBr$_{3}$ perovskite can be studied in large-scale MLFF MD
simulations. Without large-scale MD there is no guarantee to obtain the converged power spectrum when working with
VACF methods. The CsPbBr$_{3}$ perovskite has shown to form a phonon liquid
when its dynamics are projected onto harmonic eigenmodes. Only the acoustic modes close to
the $\Gamma$ point show well-behaved power spectra in the sense of 
inequality~(\ref{PeakCriterion}). The rattling motion was identified to be a nearly dispersionless 
movement of the Cs$^{+}$ cations at an effective frequency of 0.8~THz within the accessible space formed
by the PbBr$_{6}$ octahedra. Last, the dynamic stabilization is caused by collective octahedral tilting modes
only involving displacements of Br atoms.

\section{Code Availability} \label{codeavail} 
Algorithms were implemented for computing
the (P)VACF in large supercells
with long MD trajectories. The open source analysis code:
\textit{Dynamic Solids Large Ensemble Analysis Package} (DSLEAP) as well as a manual
can be downloaded from: 
\href{https://github.com/Jonathan271828/DSLEAP}{GitHub}.

\section{Acknowledgements} The authors would like to thank Ryosuke Jinnouchi and Georg Kresse for
stimulating discussions. We thank Max Rang for providing valuable feedback on the manuscript. 
We acknowledge funding by the Austrian Science Fund (FWF): P 30316-N27.
Computations were partly performed on the Vienna Scientific Cluster    
VSC3. This work was sponsored by NWO Domain Sience for the use of supercomputing facilities.

\appendix

\section{The Harmonic Approximation}\label{HarmonicPhonon}
Using the notation introduced in section \ref{NotationSec} and
defining the displacement from the equilibrium position as $\mathbf{u}_s(\mathbf{n},t)$,
for atom $s$ in unit-cell $\mathbf{n}$.
The harmonic approach uses a quadratic potential to approximate the true
potential, that is defined by the model describing the atomic interactions.
The Lagrangian of the harmonic crystal is given
by\cite{Landau:StatPhys1}
\begin{align}
        L=&\frac{1}{2}\sum_{\mathbf{n},s}m_{s}\dot{\mathbf{u}}^{2}_{s}(\mathbf{n},t)-\nonumber
        \\                                                                     
        &\frac{1}{2}\sum_{\mathbf{n},\mathbf{n}'}\sum_{s,s'}\mathbf{u}^{T}_{s'}(\mathbf{n}',t)
        \mathbf{\Lambda}_{s,s'}(\mathbf{n}-\mathbf{n'})\mathbf{u}_{s}(\mathbf{n},t),
   \label{Harm2}
\end{align} 
\noindent where
$\mathbf{\Lambda}_{s,s'}(\mathbf{n}-\mathbf{n'})$ is a $3\times3$ matrix
describing the coupling strength between a pair of atoms $s$ and $s'$
located in unit cells $\mathbf{n}$ and $\mathbf{n}'$. Because the
coupling of the atoms only depends on their relative positions the argument of
$\mathbf{\Lambda}_{s,s'}$ is $\mathbf{n}-\mathbf{n'}$
\cite{Landau:StatPhys1}. The velocities of the atoms are
$\dot{\mathbf{u}}_{s}(\mathbf{n},t)$ and the masses of the atoms are            
given by $m_{s}$. From the Lagrangian the equations of motions
can be obtained

\begin{equation}
	m_{s}\ddot{\mathbf{u}}_{s}(\mathbf{n},t) = -\sum_{\mathbf{n}',s'}
						    \mathbf{\Lambda}_{s,s'}(\mathbf{n}-\mathbf{n}')
						    \mathbf{u}_{s'}(\mathbf{n}',t),
	\label{EquMotions}
\end{equation}
\noindent
where $\ddot{\mathbf{u}}_{s}(\mathbf{n},t)$ is the acceleration of atom $s$ in unit-cell
$\mathbf{n}$. A monochromatic plane wave Ansatz for the atomic displacements
\begin{equation}
	\mathbf{u}_{s}(\mathbf{n},t)=\frac{1}{\sqrt{m_{s}}}\mathbf{e}_{s}(\mathbf{q})
		    e^{i[\mathbf{q}\mathbf{r}_{s}(\mathbf{n},t)-\omega t]} 
	\label{Harm2c}
\end{equation}
is used, where $\mathbf{e}_{s}(\mathbf{q})$ is the phonon
polarization vector. Note that the polarization vector $\mathbf{e}_{s}(\mathbf{q}$
only depends on the atomic index $s$ and is therefore the same for 
same atom types in different unit-cells. Inserting the plane wave Ansatz (\ref{Harm2c})
into the equation of motion (\ref{EquMotions}) results in
\begin{align}
	\omega^{2}\sqrt{m_{s}}\mathbf{e}_{s}(\mathbf{q})&e^{i\mathbf{q}\mathbf{r}_{s}
	                     (\mathbf{n},t)} =\\ \nonumber 
			&\sum_{\mathbf{n}',s'}\mathbf{\Lambda}_{s,s'}(\mathbf{n}-\mathbf{n}')
			\frac{1}{\sqrt{m_{s'}}}\mathbf{e}_{s'}(\mathbf{q})
			          e^{i\mathbf{q}\mathbf{r}_{s'}(\mathbf{n}',t)}.
	\label{Harmonic3}
\end{align}
\noindent
Dividing by $e^{i\mathbf{q}\mathbf{r}_{s}(\mathbf{n},t)}$ and carrying out the summation over $\mathbf{n}'$
yields 
\begin{equation}
	\omega^{2}\sqrt{m_{s}}\mathbf{e}_{s}(\mathbf{q})=\sum_{s'}\mathbf{\Lambda}_{s,s'}(\mathbf{q})
							 \frac{1}{\sqrt{m_{s'}}}\mathbf{e}_{s'}(\mathbf{q}),
	\label{Harmonic4}
\end{equation}
\noindent
where 
\begin{equation}
	\mathbf{\Lambda}_{s,s'}(\mathbf{q})=\sum_{\mathbf{n'}}\mathbf{\Lambda}_{s,s'}(\mathbf{n}-\mathbf{n}')
	                                    e^{i\mathbf{q}[\mathbf{r}_{s'}(\mathbf{n}',t)-
					    \mathbf{r}_{s}(\mathbf{n},t)]}.
	\label{Harmonic5}
\end{equation}
\noindent
Equation \ref{Harmonic4} is now divided by $\sqrt{m_{s}}$ and multiplied from the right by 
$e_{s'}(\mathbf{q})$ to give
\begin{equation}
	  \mathbf{\Lambda}_{s,s'}(\mathbf{q})     
			\frac{1}{\sqrt{m_{s}m_{s'}}}-\omega^{2}\delta_{s,s'} = 0
	\label{Harmonic6}
\end{equation}
\noindent
To obtain the solutions of this algebraic equation the determinant is taken 
\begin{equation}
	\text{det}|\mathbf{\Lambda}_{s,s'}(\mathbf{q}) 
                          	\frac{1}{\sqrt{m_{s}m_{s'}}}-\omega^{2}\delta_{s,s'}|=0,
	\label{Harmonic7}
\end{equation}
\noindent
to obtain the $3\nu$ eigenvalues $\omega_{\alpha}(\mathbf{q})$, known as the dispersion relation.
This implies that the eigenvectors must be indexed by the branch index to uniquely define them,
giving $\mathbf{e}_{s,\alpha}(\mathbf{q})$.

The harmonic approximation decomposes the real crystal into a non-interacting
phonon gas with phonon frequencies given by the dispersion relation
$\omega_{\alpha}(\mathbf{q})$.

\end{document}